\documentclass[10pt,journal,compsoc]{IEEEtran}
\newif\ifpeerreview

\peerreviewfalse

\usepackage[nocompress]{cite}

\usepackage{times}
\usepackage{epsfig}
\usepackage{graphicx}
\usepackage{amsmath}
\usepackage{amssymb}
\usepackage{amsmath}
\usepackage{cite}
\usepackage{url}
\usepackage{lipsum}
\usepackage{graphicx}
\usepackage[T1]{fontenc} 
\usepackage[english]{babel}
\usepackage[utf8x]{inputenc}
\usepackage{amsmath,amsthm}
\usepackage{color}
\usepackage[cmintegrals]{newtxmath}
\usepackage{bm} 
\usepackage{algorithm}
\usepackage{algorithmic}

\usepackage{subfig}
\usepackage{array}
\usepackage{booktabs}
\usepackage{multirow}
\usepackage{adjustbox}

\usepackage[switch]{lineno}

\title{Distance Surface for Event-Based Optical Flow}

\author{Mohammed~Almatrafi,~\IEEEmembership{Member,~IEEE,}
        ~Raymond~Baldwin,
        ~Kiyoharu~Aizawa,~\IEEEmembership{Fellow,~IEEE,}
        and ~Keigo~Hirakawa,~\IEEEmembership{Senior~Member,~IEEE}
\IEEEcompsocitemizethanks{\IEEEcompsocthanksitem M. Almatrafi is with the Department
of Electronic and Communication Engineering, Umm Al-Qura University, Al-Lith,
Saudi Arabia, 28434. \protect\\
E-mail: mmmatrafi@uqu.edu.sa.
\IEEEcompsocthanksitem R. Baldwin is with the Department of Electrical and Computer Engineering, University of Dayton, Dayton, US, 45469. \protect \\
Email: baldwinr2@udayton.edu.
\IEEEcompsocthanksitem K.Aizawa is with the , University of Tokyo, Tokyo 113-8656, Japan. \protect\\
Email: aizawa@hal.t.u-tokyo.ac.jp.
\IEEEcompsocthanksitem K. Hirakawa is with the Department of Electrical and Computer Engineering, University of Dayton, Dayton, US, 45469. \protect \\
Email: khirakawa1@udayton.edu.}
}

\begin{document}

\IEEEtitleabstractindextext{%
\begin{abstract} 

We propose DistSurf-OF, a novel optical flow method for neuromorphic cameras. Neuromorphic cameras (or event detection cameras) are an emerging sensor modality that makes use of dynamic vision sensors (DVS) to report asynchronously the log-intensity changes (called ``events'') exceeding a predefined threshold at each pixel. In absence of the intensity value at each pixel location, we introduce a notion of ``distance surface''---the distance transform computed from the detected events---as a proxy for object texture. The distance surface is then used as an input to the intensity-based optical flow methods to recover the two dimensional pixel motion. Real sensor experiments verify that the proposed DistSurf-OF accurately estimates the angle and speed of each events.

\end{abstract} 

}

\maketitle
\thispagestyle{empty}

\IEEEraisesectionheading{
\section{Introduction}\label{sc:intro}
}

\emph{Optical flow} refers to the task of estimating the apparent motion in a visual scene. It has been a major topic of research in computer vision for the past few decades due to the significant role it plays in various machine vision applications, including navigation \cite{mueggler2018continuous,gallego2018unifying,gallego2019focus,zhu2018unsupervised}, segmentation \cite{stoffregen2018simultaneous,stoffregen2019event}, image registration \cite{bardow2016simultaneous}, tracking \cite{kim2014simultaneous,clady2017motion,gehrig2018asynchronous}, and motion analysis \cite{chen2019fast}. While remarkable progress have been made since the original concepts were introduced by Horn-Schunck~\cite{horn1981determining} and Lucas-Kanade~\cite{lucas1981iterative}, optical flow in the presence of fast motion and occlusions remains a major challenge today \cite{barron1994performance,baker2011database,sun2014quantitative}.

\begin{figure}
\begin{tabular}{@{}c@{~}c@{}}
\includegraphics[width=.24\textwidth]{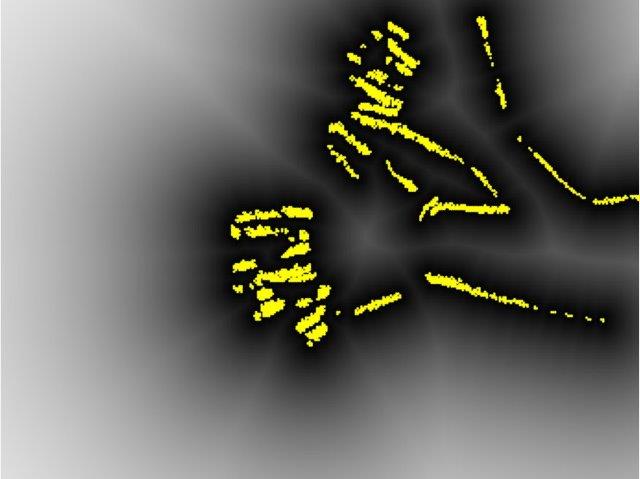}&
\includegraphics[width=.24\textwidth]{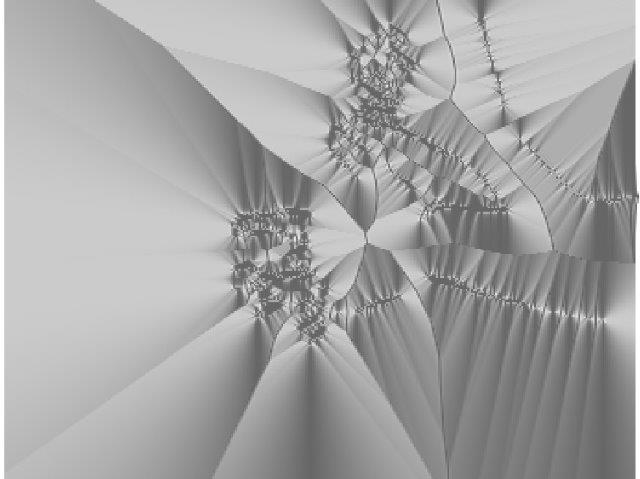}\\
(a) Distance surface $D(\bm{X},k)$ &
(b) Spatial derivative $\frac{\partial}{\partial x}D(\bm{X},k)$\\
\includegraphics[width=.24\textwidth]{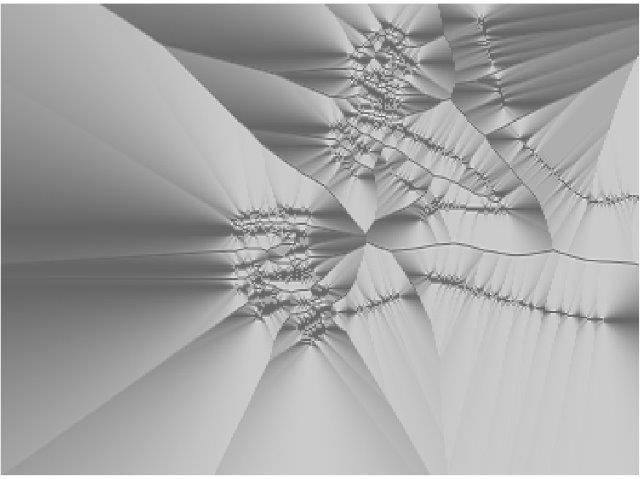}&
\includegraphics[width=.24\textwidth]{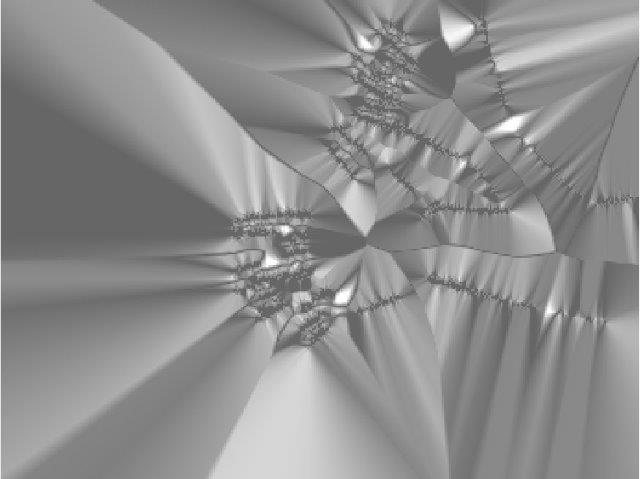}\\
(c) Spatial derivative $\frac{\partial}{\partial y}D(\bm{X},k)$&
(d) Temporal derivative $\frac{\partial}{\partial t}D(\bm{X},k)$\\
\end{tabular}
 \caption{Spatial/temporal derivatives of the proposed distance surface on \emph{hand} sequence. Yellow pixels denote denoised events.
 }
 \label{fig:deriv}
\end{figure}

In recent years, neuromorphic cameras have gained popularity in applications that require cameras to handle high dynamic range and fast scene motion scenes. Unlike the conventional active pixel sensor (APS) that records an image intensity at a (slow) synchronous frame-rate, DVS in neuromorphic cameras asynchronously reports spikes called ``events'' when the log-brightness change exceeds a fixed threshold. Since these events only occur at object edges, they are very sparse. See Figure~\ref{fig:deriv}(a). DVS represents a significant reduction in memory storage and computational cost, increase in temporal resolution (+800kHz), higher dynamic range (+120dB), and lower latency (in the order of microseconds). Thus, neuromorphic cameras have the potential to improve the performance of optical flow methods, which are currently limited by the slow frame rate of the conventional camera's hardware. The main challenges to working with neuromorphic cameras, however, is the lack of the notion of pixel intensity, which renders conventional image processing and computer vision tools useless.

In this work, we propose {\bf DistSurf-OF}, a novel DVS-based optical flow method that is robust to complex pixel motion vectors and scenes. We achieve this by introducing a novel notion of ``distance surface,'' designed to corroborate pixel velocity from multiple edge pixels of varying edge orientations. We interpret the distance surface as a proxy for pixel intensity values in conventional cameras and treat its spatial derivatives as the ``object textures'' of non-edge pixels. This disambiguates the pixel motion and its temporal derivative as the encoding of the texture changes over time. See Figure~\ref{fig:deriv}(b-d). The computed distance surface derivatives are then used as an input to the standard optical flow methods to recover the two dimensional pixel motion. 

The main contributions of this paper are as follows:
\begin{itemize}
    \item {\bf Distance Surface:} We assign an intensity value to each pixel based on the proximity to the spatially closest detected event pixel. These values represent the object shapes by the relative positions of their edges, which satisfy the optical flow equations.
    \item {\bf DistSurf-OF:} We recover the pixel motion field from the spatial-temporal gradients of the distance surface. The computed motion field draws on multiple events corresponding to multiple edge orientations, improving the robustness to motion and scene complexity.
    \item {\bf Noise Robustness Study:} We employ event denoising to improve optical flow performance. We also analyze its positive impact on existing DVS-optical flow methods.
    \item {\bf DVSMOTION20:} We present a new optical flow dataset of complex scenes and camera motion. Ground truth motion can be inferred from the rotational camera motion measured by the inertial measurement unit.
\end{itemize}

\begin{figure*}
  \centering
    \subfloat[APS Data]{\includegraphics[width=.25\textwidth]{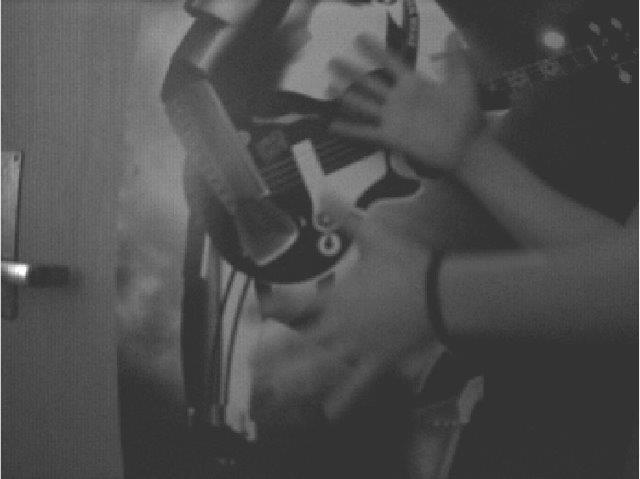}}~
    \subfloat[DVS Data]{\includegraphics[width=.25\textwidth]{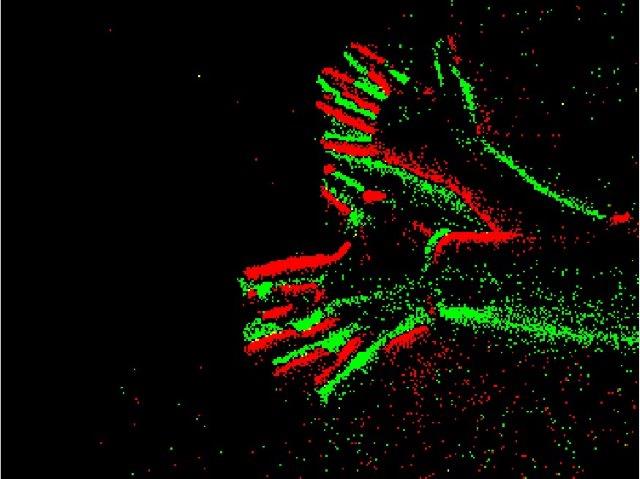}}~
    \subfloat[APS vs DVS as function of time]{\includegraphics[width=.25\textwidth]{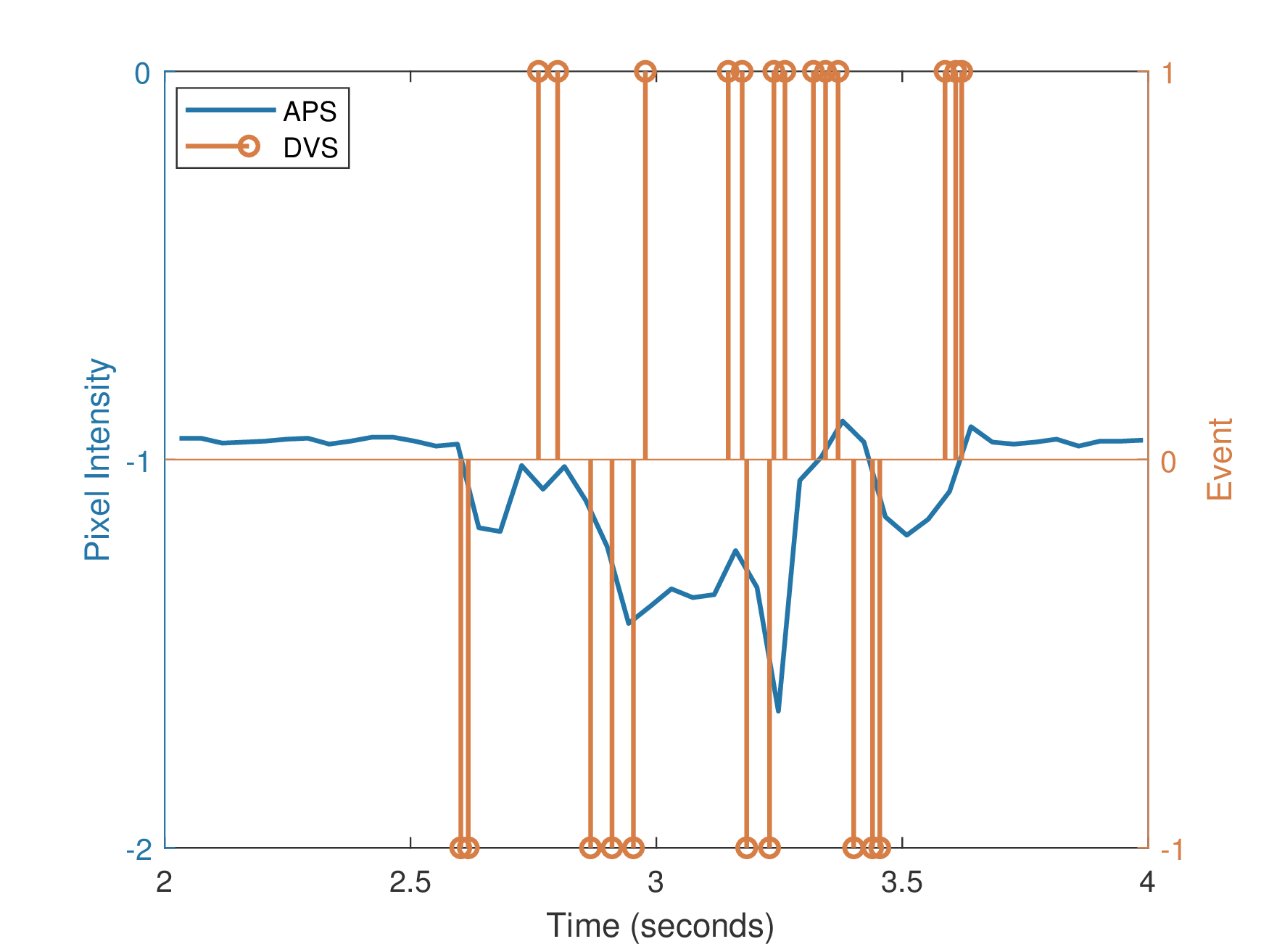}}~
    \subfloat[Spatial-temporal data]{\includegraphics[width=.25\textwidth]{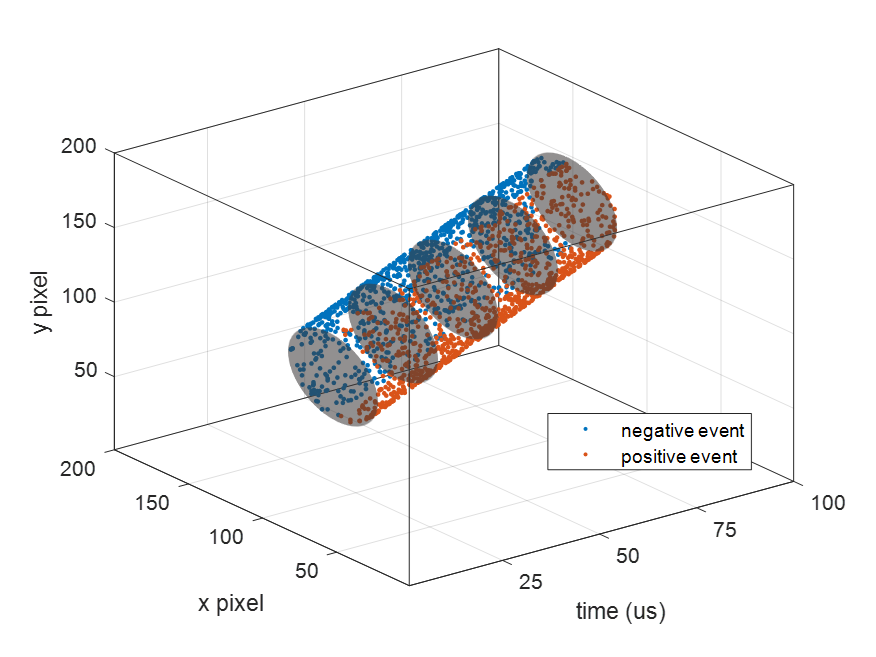}}\\
  \caption{Example outputs from DAViS sensor, which has APS and DVS readout circuits sharing the same photodiode. (a) APS's synchronous outputs record pixel intensities. (b) DVS's asynchronous events report positive (green) and negative (red) changes in a pixel intensity. (c) Superposition of DVS and APS outputs of one particular pixel. (d) An example of a dark disk moving quickly across the scene. APS synchronously records the disk every 25ms (at 40Hz). DVS operates at 1MHz, tracking the disk’s movement in between APS frames.}  
  \label{fig:APS_DVS}
\end{figure*}

The remainder of this paper is organized as follows. In Section~\ref{sec:DVS}, we briefly review the requisite background materials. In Section~\ref{sec:proposed}, we then propose the proposed distance surface-based optical flow method for DVS cameras. Section~\ref{sec:denoising} outlines methods to increase algorithm robustness. We describe DVSMOTION20 and present the real-data experimental results in Section~\ref{sec:experiments} before making the concluding remarks in Section~\ref{sec:conclusion}.



\section{Background and Related Work}\label{sec:DVS}

\subsection{Frame-Based Optical Flow} \label{OF}
Let $I:\mathbb{Z}^2 \times \mathbb{R}\to \mathbb{R}$ denote an intensity video, 
where $I(\bm{X},t)$ is the pixel radiance at pixel $\bm{X}=(x,y)^T\in\mathbb{Z}^2$ of a video frame at time $t\in\mathbb{R}$. Known as the ``brightness constancy assumption,'' optical flow is derived from the hypothesis that the the pixel intensities of translated objects remain constant over time \cite{horn1981determining}:
\begin{equation}\label{eq:constancy}
 I(\bm{X}+\Delta\bm{ X},t+\Delta t) = I(\bm{X},t),
\end{equation}
where $\Delta\bm{X}=(\Delta x,\Delta y)$ and $\Delta t$ denote spatial and temporal translations, respectively. Assuming that such translations are small,  \eqref{eq:constancy} is expanded via first-order Taylor series to give rise to the well-known ``optical flow equation'':
\begin{align} \label{OFeq}
 \nabla I(\bm{X},t)\bm{V}(\bm{X},t) +   I_t(\bm{X},t)  \approx 0,
  \end{align} 
where $\nabla I(\bm{X},t) = (\frac{\partial}{\partial x}I(\bm{X},t),\frac{\partial}{\partial y}I(\bm{X},t))$  and $I_t(\bm{X},t)=\frac{\partial}{\partial t}I(\bm{X},t)$ are the spatial gradient and temporal derivative, respectively. The goal of the optical flow task is to estimate the two dimensional pixel motion field $\bm{V}(\bm{X},t)=(\frac{\Delta x}{\Delta t}\frac{\Delta y}{\Delta t})^T$. 

The pixel motion field cannot be estimated directly from the under-determined system of equations in \eqref{OFeq} because the component of $\bm{V}$ parallel to the edges lives in the nullspace of $\nabla I(\bm{X},t)$. We overcome this issue---commonly referred to as the ``aperture problem''---by imposing additional constraints. An example of such constraint is the flow magnitude minimization. However, its solution is a motion perpendicular in direction to the edge orientation---a phenomenon often referred to as the ``normal flow''---which does not necessarily represent the actual two dimensional motion of the object in general. Overcoming the normal flow problem requires diversifying the gradient orientation $\nabla I(\bm{X},t)$ by incorporating multiple pixels. The ``local spatial consistency'' constraint proposed by Lucas-Kanade helps overcome noise and variations by requiring $\bm{V}(\bm{X},t)$ to be the same within a spatial neighborhood~\cite{lucas1981iterative}. Similarly, Horn-Schunck (HS) introduced  ``global spatial constancy'' criteria promoting smoothness of $\bm{V}(\bm{X},t)$ globally by adding a quadratic penalty to \eqref{OFeq} as a regularization term\cite{horn1981determining}. More recently developed methods improve upon these classical optical flow methods to yield state-of-the-art performance \cite{sun2010secrets} by modifying penalty terms~\cite{barron1994performance,bruhn2005lucas,baker2004lucas,sun2014quantitative} and leveraging phase~\cite{fleet1990computation,fleet1993stability,gautama2002phase} and block~\cite{lowe2004distinctive,kroeger2016fast} correlations.

Conventional optical flow methods applied to an intensity video sequence with a relatively slow frame rate (e.g.~30 or 60 frames per second) often leads to unreliable optical flow. Instability stems from large translations $\Delta \bm{X}$ and $\Delta t$ in the presence of fast motion~\cite{sun2014quantitative,bao2014fast}, which invalidate the first order Taylor series approximation in \eqref{OFeq}. Similarly, variations in illumination conditions and occlusion cause changes over time in the intensity values of spatially translated features, negating the brightness constancy assumption in \eqref{eq:constancy}. Current solutions to deal with these challenges include spatial pyramids~\cite{memin1998multigrid,memin2002hierarchical,black1996robust,bao2014fast}, local layering~\cite{Sun_2014_CVPR}, and robust penalty functions such as generalized Charbonnier~\cite{sun2010secrets}, Charbonnier~\cite{bruhn2005lucas}, and  Lorentzian~\cite{black1996robust}. Additionally, the reliability of optical flow can be also enhanced by outlier removals~\cite{wedel2009improved}, texture decomposition~\cite{wedel2008duality,wedel2009improved}, and smoothing filters~\cite{barron1994performance,bruhn2005lucas} to preserve the brightness assumption. 

\subsection{Dynamic Vision Sensor (DVS)}




Instead of recording an image intensity at a synchronous frame-rate, neuromorphic cameras record asynchronous positive and negative spikes called events. These events are generated when the log-brightness change exceeds a fixed threshold $\ell$: 
\begin{align}\label{eq:events}
\begin{split}
t_{i+1}(\bm{X})=&\arg\min_{t}
\left\{t>t_i\Bigg| \left|\log\left(\frac{I(\bm{X},t)}{I(\bm{X},t_{i})}\right)\right| > \ell \right\}\\
p_{i+1}(\bm{X})=&\operatorname{sign}\left(I(\bm{X},t_{i+1})-I(\bm{X},t_{i})\right),
\end{split}
\end{align} 
where $t_{i+1}(\bm{X})$ denotes the time of event occurrence, whose accuracy is in the order of microseconds; and $p_{i+1}(\bm{X})$ is the polarity, indicating whether the intensity change is darker (-1) or brighter (+1). 
The sparsity of the events reduces throughput and memory storage considerably, enabling high temporal resolution and low latency. It has been widely speculated that such characteristics can help overcome the limitations of conventional frame-based optical flow methods. Specifically,  the microseconds resolution implies $\Delta \bm{X}$ and $\Delta t$ are small, better preserving the validity of the Taylor series expansions in \eqref{OFeq} and minimizing the risks of occlusion, even in the presence of fast motion. 

Optical flow for DVS is a task of determining the velocities of the pixels that generated the observed events. Because of the fact that DVS outputs lack the notion of pixel intensity (and brightness constancy assumption in \eqref{eq:constancy} is largely invalid), optical flow requires an entirely new approach. Prior efforts were aimed at creating a proxy to image intensity (via accumulation of events over a temporal window)~\cite{benosman2012asynchronous}, the spatial gradient images (via central difference)~\cite{rueckauer2016evaluation}, and the temporal derivative (via second order backward difference)~\cite{brosch2015event}. 
Early DVS-specific optical flow approaches include local plane fitting that infers the pixel motion by fitting spatial-temporal manifold to the events based on their time stamps~\cite{benosman2014event,rueckauer2016evaluation,brosch2015event}, edge orientations estimation~\cite{delbruck2008frame,lee2012touchless}, and block matching~\cite{liu2017block,liu2018abmof}. 
Recent event-based techniques simultaneously estimate optical flow along with other machine vision tasks, such as intensity estimation~\cite{bardow2016simultaneous}, depth along with motion reconstruction~\cite{gallego2018unifying}, contrast maximization~\cite{gallego2018unifying,stoffregen2019event}, and segmentation~\cite{stoffregen2018simultaneous,stoffregen2019event}. 
Appealing to recent successes of machine learning approaches to optical flow~\cite{jason2016back,ilg2017flownet,meister2018unflow}, learning-based optical flow methods for neuromorphic cameras have recently been proposed~\cite{zhu2018ev,ye2018unsupervised,zhu2018unsupervised,paredes2019unsupervised}. Finally, the method in~\cite{almatrafi2019davis} exploits the high spatial fidelity of APS and temporal fidelity of DVS to compute spatial temporal derivatives for hybrid hardware called DAViS\cite{brandli2014240}. Benchmarking datasets for DVS-based optical flow and real-time performance evaluation platform have helped accelerate the progress of research in optical flow for neuromorphic cameras~\cite{rueckauer2016evaluation,zhu2018ev,almatrafi2019davis}.

\section{Proposed: DistSurf Optical Flow} \label{sec:proposed}

\begin{figure}
\centering
\includegraphics[width=.5\textwidth]{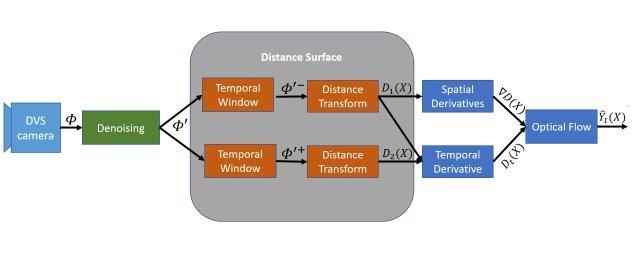}
\caption{System diagram for DistSurf-OF.}
\label{fig:system}
\end{figure}

\begin{figure}
    \centering
\begin{tabular}{@{}c@{~}c@{}}
\includegraphics[width=.35\textwidth,height=.35\textwidth]{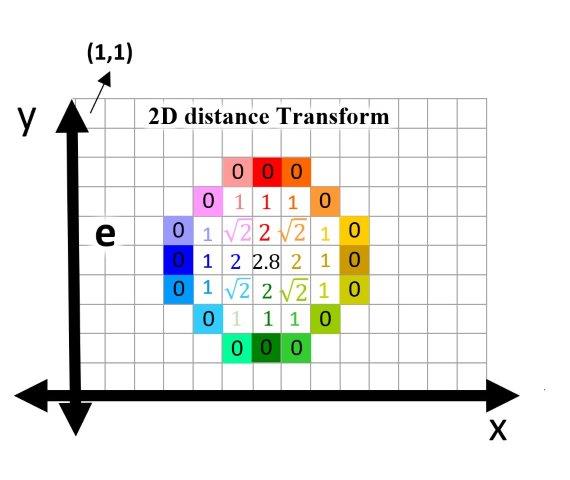}\\
\end{tabular}
    \caption{An example of a distance transform. Solid colored pixels indicate event/edge pixels. Numbers indicate the minimum distance between a pixel and an edge. The color of the numbers indicate the closest edge pixel.}
    \label{fig:direction}
\end{figure}

We propose a new DVS-based optical flow method designed to estimate the apparent motion in the scene at the edge pixels. See Figure~\ref{fig:system} for an overview. The underlying assumption is that the pixel spatial velocities of a rigid- or semirigid-body object are slowly varying. As such, the pixel velocity of a pixel internal to a semiregid-body object can be inferred from the edge pixels surrounding it. We propose a novel notion of ``distance surface'' as a way to leverage multiple edges of a semirigid-bodied object and as a proxy for object textures---incorporating multiple edges avoids the pitfalls of normal flow that many existing DVS-based optical flow algorithms suffer from.


\subsection{Distance Surface} \label{sec:distsurf}

We are interested in recovering the two dimensional motion field $\bm{V}(\bm{X},t)$ from events detected immediately before and immediately after the time $t\in\mathbb{R}$. Define $\Phi\subset\mathbb{Z}^2$ as a set of pixel indices corresponding to detected events that occurred within a temporal window $[t-\Delta t,t)$, as follows:
\begin{align}\label{eq:edge}
\Phi =& \{\bm{X} \in \mathbb{Z}^2 | \exists i\ni t-\Delta t  \leq t_{i}(\bm{X}) < t \},
\end{align}
where $t_{i}(\bm{X})$ denotes the time-stamps of the events as described in Section~\ref{sec:DVS}. 
Let $d:\mathbb{Z}^2\times \mathbb{Z}^2\to \mathbb{R}$ be a distance measure function, with $d(\bm{X},\bm{Y})$ representing the spatial distance between pixels $\bm{X}=(X_1,X_2)^T\in\mathbb{Z}^2$ and $\bm{Y}=(Y_1,Y_2)^T\in\mathbb{Z}^2$. In this work, we consider the  $L^2$ norm taking the form:
\begin{align}\label{eq:WL2}
d(\bm{X},\bm{Y}) &= \sqrt{(\bm{X}-\bm{Y})^T(\bm{X}-\bm{Y})}.
\end{align}
Then, a distance transform converts the sets $\Phi\in\mathbb{Z}$ to gray-level image $D:\mathbb{Z}^2\to\mathbb{R}$ by:
\begin{align}\label{eq:distsurf}
D(\bm{X}) =& \min_{\bm{Y} \in \Phi} d(\bm{X},\bm{Y})=d\left(\bm{X},\widehat{\bm{Y}}(\bm{X})\right)
\end{align}
with corresponding indexes $\widehat{\bm{Y}}:\mathbb{Z}^2\to\mathbb{Z}^2$ as edge pixels deemed ``closest'' to $\bm{X}$:
\begin{align}\label{eq:Y}
    \widehat{\bm{Y}}(\bm{X})=&\arg\min_{\bm{Y}\in\Phi} d(\bm{X},\bm{Y}).
\end{align}
An example of distance transform is illustrated in Figure~\ref{fig:direction}. It is clear that $D(\bm{X})=0$ when $\bm{X} \in \Phi$. At a non-edge pixel $\bm{X} \notin \Phi$, the pixel intensity $D(\bm{X})>0$ represents the distance to the nearest edge pixel $\widehat{\bm{Y}}(\bm{X})\in\Phi$.

We propose to use the gray-level image $D(\bm{X})$ computed from DVS as a proxy for intensity value in APS images---a notion we hereafter refer to as ``distance surface.'' That is, we treat the distance surface $D(\bm{X})$ as object textures on pixels that are internal to an object, away from edges. While a full justification for using distance surface in the context of optical flow is provided in Section~\ref{sec:opticalflow}, we can already see in Figure~\ref{fig:direction} some of the reasons that the notion of distance surface is well suited for the optical flow task---although $D(\bm{X})$ is a smooth function of $\bm{X}$, the spatial gradient of $D(\bm{X})$ is influenced by the selection of different edge pixels $\widehat{\bm{Y}}(\bm{X})\in\Phi$ closest to pixel $\bm{X}$ internal to the rigid-bodied object. As such, an optical flow computed from the distance surface implicitly incorporates multiple edge pixels (of multiple edge orientations) to establish an object motion, so as to overcome the normal flow problem that hampers many optical flow algorithms. 

\begin{figure}
\centering
\includegraphics[width=.3\textwidth]{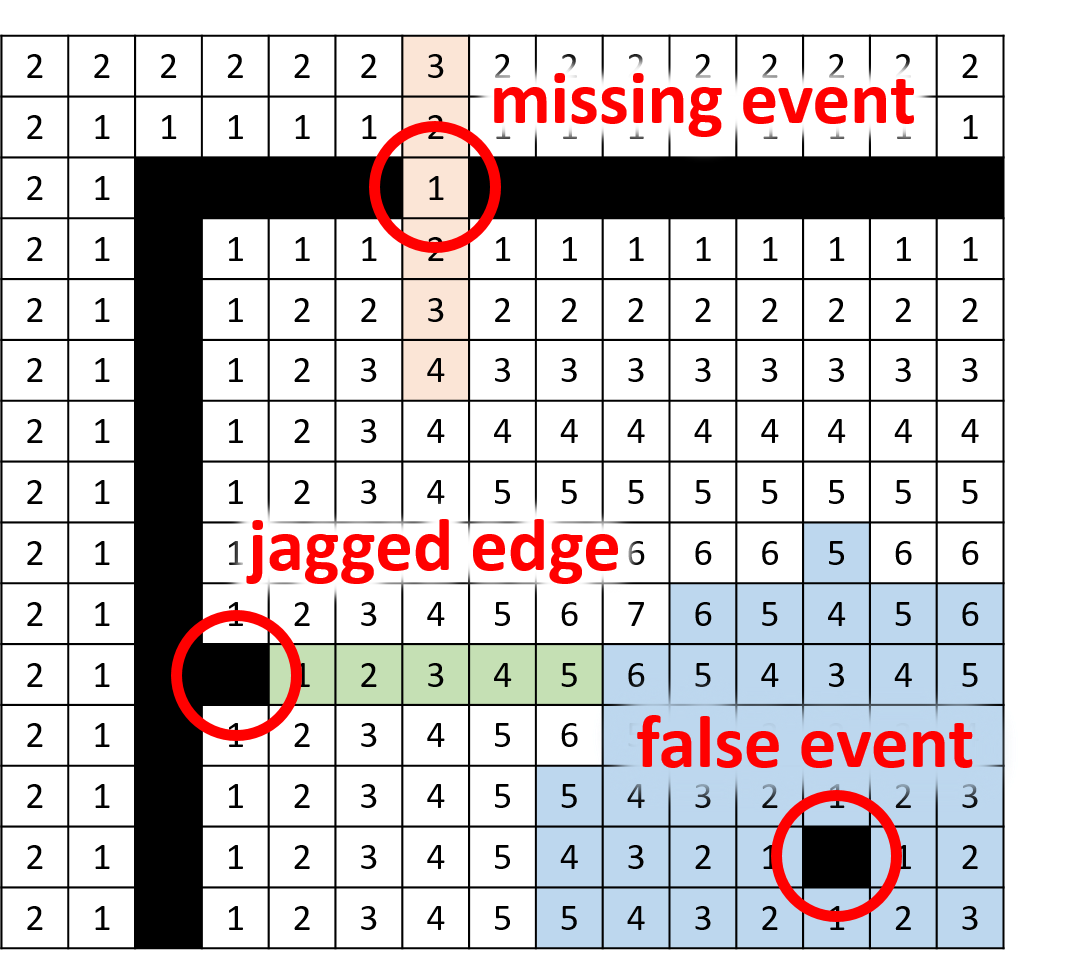}
\caption{An illustration of distance transform (shown using $L^1$ distance measure function rather than $L^2$ used in our algorithm because it is easier to follow). Black pixels denote detected events. Distance transform is robust to ``missing'' events and ``jagged edges''---as evidenced by orange- and green-shaded pixels, the affected pixels are negligibly different from the desired distance transform values. By contrast, false positive events deteriorate a large number of the pixels (shaded in blue) by a large margin.}
\label{fig:DT}
\end{figure}

Our work in distance surface can be interpreted as an alternative to the ``time surface''~\cite{lagorce2017hots}---time surface, too, can be interpreted as a type of a distance transform (in the temporal domain):
\begin{align}
    T(\bm{X})=\min_{t_i(\bm{X})<t} \|t-t_i(\bm{X})\|.
\end{align}
Using time surface as a proxy for pixel intensity values in APS images has been proven useful in pattern recognition applications~\cite{alzugaray2018asynchronous,sironi2018hats}. However, the brightness constancy assumption is invalid for $T(\bm{X},k)$, making it a poor choice for optical flow. 

Finally, note that the definition of distance transform in \eqref{eq:distsurf} makes no distinction of the polarity of the DVS generated events, as our goal is to recover the edges of the rigid-bodied objects surrounding the internal pixels $\bm{X}$---an intuition confirmed via empirical experiments. Since the polarity of the events occurring at the object boundary is an encoding of the contrast between the foreground and the background, coupling the pixel motion estimation to the background texture properties only destabilized our optical flow method.

\subsection{Distance Surface Optical Flow Equation} 
\label{sec:opticalflow}

\begin{figure}
    \centering
    \includegraphics[width=.3\textwidth]{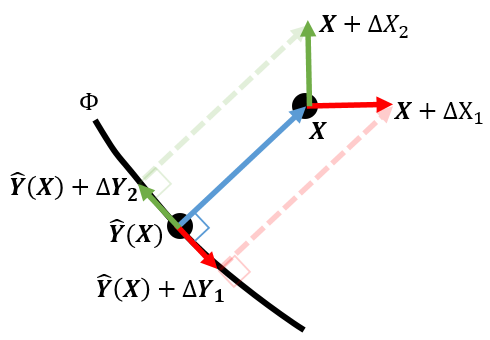}
    \caption{Event $\widehat{\bm{Y}}(\bm{X})\in\Phi$ is the point on the edge manifold $\Phi$ closest to pixel location $\bm{X}$. At $\widehat{\bm{Y}}(\bm{X})$, the gradient vectors (red and green arrows) are orthogonal to $\bm{X}-\widehat{\bm{Y}}(\bm{X})$ (blue arrow). Thus, $\bm{X}-\widehat{\bm{Y}}(\bm{X})$ lives in the nullspace of the Jacobian matrix in \eqref{eq:J} (whose column vectors are the gradient vectors of $\widehat{\bm{Y}}(\bm{X})$).}
\label{fig:proof}
\end{figure}

The distance surface in \eqref{eq:distsurf} is very well matched for the DVS optical flow task. To understand why this is the case, we begin by first proving that the distance surface satisfies the optical flow equation in \eqref{OFeq}. Taking a spatial gradient of the distance surface, we have the following relation:
\begin{align}\label{eq:J}
\begin{split}
    \nabla D(\bm{X})
    =&
    \frac{\left(\bm{X}-\widehat{\bm{Y}}(\bm{X})\right)^T}{d\left(\bm{X},\widehat{\bm{Y}}(\bm{X})\right)} - \frac{\left(\bm{X}-\widehat{\bm{Y}}(\bm{X})\right)^T\bm{J}_{\widehat{\bm{Y}}}\left(\bm{X}\right)}{d\left(\bm{X},\widehat{\bm{Y}}(\bm{X})\right)}\\
    =&
    \frac{\left(\bm{X}-\widehat{\bm{Y}}(\bm{X})\right)^T}{d\left(\bm{X},\widehat{\bm{Y}}(\bm{X})\right)},    
\end{split}
\end{align}
where above, we used the fact that the vector $\bm{X}-\widehat{\bm{Y}}(\bm{X})$ is in the nullspace of the Jacobian matrix $\bm{J}_{\widehat{\bm{Y}}}\in\mathbb{R}^{2\times 2}$ in the last step (see Figure \ref{fig:proof}). Similarly, consider the temporal derivative of the distance surface:
\begin{align}
    D_t\left(\bm{X}\right):=\frac{\partial}{\partial t}D\left(\bm{X}\right)=&-
    \frac{\left(\bm{X}-\widehat{\bm{Y}}(\bm{X})\right)^T}{d\left(\bm{X},\widehat{\bm{Y}}(\bm{X})\right)} \widehat{\bm{Y}}_t(\bm{X}),
\end{align}
where $\widehat{\bm{Y}}_t(\bm{X}):=\frac{\partial}{\partial t}\widehat{\bm{Y}}(\bm{X})$ is the event pixel velocity we are after. Thus, we arrive at a new optical flow equation for distance surface:
\begin{align}\label{eq:OFeq_distsurf}
    \nabla D\left(\bm{X}\right)\widehat{\bm{Y}}_t(\bm{X})+D_t\left(\bm{X}\right)=0.
\end{align}
Note that unlike the classical optical flow equation in \eqref{OFeq}, distance surface optical flow equation in \eqref{eq:OFeq_distsurf} is exact (i.e.~derived without Taylor series approximation), meaning it is robust to fast motions.

Hence, we propose to recover the event pixel velocity $\widehat{\bm{Y}}_t(\bm{X})$ by leveraging distance surface optical flow equation in \eqref{eq:OFeq_distsurf}. That is, we solve for $\widehat{\bm{Y}}_t(\bm{X})$ using classical optical flow methods. Spatial derivative $\nabla D(\bm{X})$ may be computed directly from the distance transform $D(\bm{X})$ using a spatial derivative convolution filter (the same way $\nabla I(\bm{X})$ in \eqref{OFeq} is computed in conventional optical flow methods):
\begin{align}\label{eq:spatialgradient}
    \nabla D(\bm{X})=
    \begin{bmatrix}
    \frac{\partial}{\partial X_1}D(\bm{X})\\
    \frac{\partial}{\partial X_2}D(\bm{X})
    \end{bmatrix}
=
    \begin{bmatrix}
    H_X(\bm{X})\star D(\bm{X})\\
    H_Y(\bm{X})\star D(\bm{X})
    \end{bmatrix}
\end{align}
for some horizontal and vertical spatial derivative filters $H_X,H_Y:\mathbb{Z}^2\to\mathbb{R}$. For temporal derivative, we consider difference of two distance surfaces:
\begin{align}
\begin{split}
    D_t(\bm{X})=&\frac{1}{\Delta t}\left(\left(\min_{\bm{Y}\in\Phi^+} d(\bm{X},\bm{Y})\right)-\left(\min_{\bm{Y}\in\Phi^-} d(\bm{X},\bm{Y})\right)\right),
\end{split}
\end{align}
where the sets $\Phi^-,\Phi^+\subset\mathbb{Z}^2$ correspond to a narrow temporal window immediately before and after $t\in\mathbb{R}$:
\begin{align}
\begin{split}
    \Phi^-=&\{\bm{X}\in\mathbb{Z}^2|\exists i\ni t-\Delta t\leq t_i(\bm{X})<t\}\\
    \Phi^+=&\{\bm{X}\in\mathbb{Z}^2|\exists i\ni t\leq t_i(\bm{X})<t+\Delta t\}.
\end{split}
\end{align}
Spatial and temporal gradients $\nabla D(\bm{X})$ and $D_t(\bm{X})$ may be used subsequently as the input to \emph{any} conventional optical flow method to yield an estimation of pixel velocity $\widehat{\bm{Y}}_t(\bm{X})$ at every pixel.

In our implementation, we used the classical Horn Schunck approach~\cite{horn1981determining} aimed at minimizing the global energy functional $\bm{E}_{HS}:\ell^2(\mathbb{Z}^2)\to\mathbb{R}$ given by 
\begin{align} \label{eq:HS-general}
\begin{split}
\bm{E}_{HS}\left(\widehat{\bm{Y}}_t(\bm{X})\right)
&= \sum_{\bm{X}\in\mathbb{Z}^2}  \rho\Big(\nabla D(\bm{X})\widehat{\bm{Y}}_t(\bm{X}) + D_t(\bm{X})\Big) +  \lambda~ \rho\Big(\nabla \widehat{\bm{Y}}_t(\bm{X})\Big).
\end{split}
\end{align} 
Here, $\lambda$ is the regularization parameter (set to 0.1 in this work) and $\rho(x)=\log(1+\frac{\|x\|^2}{2\sigma^2})$ denotes Lorentzian robust penalty function\cite{black1996robust} as implemented in~\cite{sun2014quantitative}. The regularization term in effect imposes spatial consistency on the estimated event pixel velocities based on neighboring events that comprise an object boundary shape. By relying on multiple events within the contour of the object, we diversify the edge orientations and reduce the risk of normal flow motion artifacts, increase robustness to noise, etc. Readers are reminded that while the method in \eqref{eq:HS-general} yields a state-of-the-art result, the proposed DistSurf-OF framework is agnostic to the choice of intensity-based optical flow method it is paired with in general.

Recalling the mapping in \eqref{eq:Y}, suppose there are multiple non-edge pixels $\{\bm{X}_1,\dots,\bm{X}_N\}$ that map back to the same edge pixel:
\begin{align}
\bm{Y}=\widehat{\bm{Y}}(\bm{X}_1)=\widehat{\bm{Y}}(\bm{X}_2)=\dots=\widehat{\bm{Y}}(\bm{X}_N).    
\end{align}
Then technically, $\left\{\widehat{\bm{Y}}_t(\bm{X}_1),\widehat{\bm{Y}}_t(\bm{X}_2),\dots,\widehat{\bm{Y}}_t(\bm{X}_N)\right\}$ are all valid estimates of the event pixel velocity $\bm{V}(\bm{Y},t)$. In our work, we chose a simplistic and computationally efficient approach. Since $\widehat{Y}(\bm{Y})=\bm{Y}$ (i.e.~an event pixel $\bm{Y}\in\Phi$ is closest to itself), we assigned the estimated distance surface velocity to the final event pixel velocity $\bm{V}(\bm{Y},t)$ as follows:
\begin{align}
\bm{V}\left(\bm{Y},t\right)=\widehat{\bm{Y}}_t(\bm{Y}).
\end{align}
However, a more sophisticated approach combining $\left\{\widehat{\bm{Y}}_t(\bm{X}_1),\widehat{\bm{Y}}_t(\bm{X}_2),\dots,\widehat{\bm{Y}}_t(\bm{X}_N)\right\}$ to yield the final estimate $\bm{V}\left(\bm{Y},t\right)$ may increase robustness to perturbations. This is left as a future topic of investigation.

\section{Robustness Analysis and Denoising}
\label{sec:denoising}
Let $\Psi$ in \eqref{eq:edge} denote the ``ideal'' event detection corresponding to the image edges. DVS suffers from considerably high noise (random events) along with the signal due to multiple factors such as electronic noise and sensor heat~\cite{padala2018noise}. As such, the set of actual observed events $\Phi$ is a perturbed version of the ideal set of events $\Psi$ in the following sense:
\begin{align}
    \Phi = \{\Psi\cap \bar{\Lambda}\} \cup \Omega,
\end{align}
where $\Omega\subset\mathbb{Z}^2$ is a set of random DVS activations (i.e.~false positives); $\Lambda\subset \Psi$ is a set of ``holes'' or missing events (i.e.~false negatives) randomly excluded from $\Psi$; and $\bar{\Lambda}$ denotes the complimentary set to $\Lambda$. \emph{What is the practical impact of computing the distance transform using $\Phi$ instead of $\Psi$?}

The distance transform in \eqref{eq:distsurf} is robust to ``holes'' in the edge pixels of the binary frame $\Psi$. To understand why this is the case, let $\bm{Y}\in\Psi$ and $\bm{Z}\in\Psi$ be the edge pixels where $d(\bm{X},\bm{Y})<d(\bm{X},\bm{Z})$---that is, $\bm{Y}$ is closer to $\bm{X}$ than $\bm{Z}$. Then by the triangle inequality of distance measure functions, we have the following relation:
\begin{align}
d(\bm{X},\bm{Z})\leq d(\bm{X},\bm{Y})+d(\bm{Y},\bm{Z}).
\end{align}
Suppose further that $\bm{Y}\in\Lambda$ is missing (i.e.~$\bm{Y}\notin\Phi$). Then the penalty for replacing $d(\bm{X},\bm{Y})$ by $d(\bm{X},\bm{Z})$ is
\begin{align}
 0< d(\bm{X},\bm{Z})-d(\bm{X},\bm{Y})\leq d(\bm{Y},\bm{Z}).
\end{align}
Owing to the fact that edge pixels occur in clusters (i.e.~$ d(\bm{Y},\bm{Z})$ is small), we conclude $d(\bm{X},\bm{Y})\approx d(\bm{X},\bm{Z})$. Thus, the distance transform $D(\bm{X},k)$ is largely invariant to  random exclusions of events $\Lambda$ in $\Psi$:
\begin{align}
    \min_{\bm{Y}\in\Psi}d(\bm{X},\bm{Y})\approx \min_{\bm{Y}\in\Psi\cap \bar{\Lambda}}d(\bm{X},\bm{Y}).
\end{align}
Same analysis applies to jagged edges, where detected events at the edges are displaced by one or two pixels. See Figure~\ref{fig:DT} for examples.

On the other hand, the distance surface is vulnerable to randomly activated events (i.e.~false events) in $\Omega$. Let us rewrite the distance transform as follows:
\begin{align}
    \min_{\bm{Y}\in\Psi\cup \Omega}d(\bm{X},\bm{Y})= \min\left\{\min_{\bm{Y}\in\Psi}d(\bm{X},\bm{Y}),\,\, \min_{\bm{Z}\in\Omega}d(\bm{X},\bm{Z})\right\}.
\end{align}
In other words, random distance transform $\min_{\bm{Z}\in\Omega}d(\bm{X},\bm{Z})$ is a source of significant degradation to the desired distance surface $\min_{\bm{Y}\in\Psi}d(\bm{X},\bm{Y})$. As such, the severity of degradation increases with the distance $d(\bm{X},\bm{Y})$. See Figure~\ref{fig:DT}. Therefore, the proposed use of distance surface would benefit from denoising randomly activated events.

\begin{figure}
\centering
\includegraphics[width=0.95\linewidth]{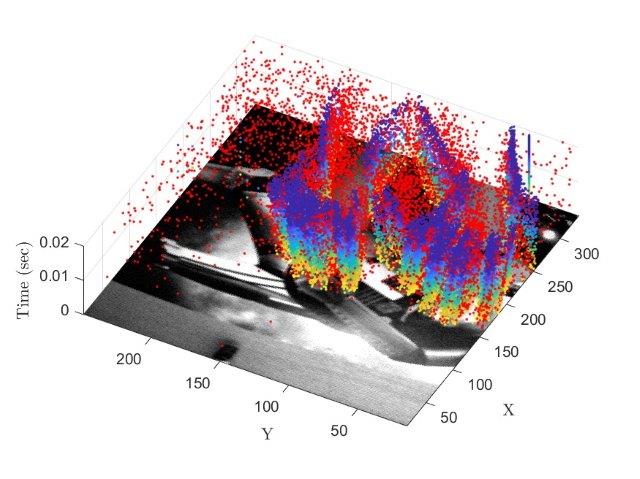}
   \caption{Example of denoising. Red points indicate BA (or randomly activated events) that we filter out. Our optical flow method is being applied to IE and TE events that comprise remainder of points.}
\label{fig:eventGen}
\end{figure}

Denoising used in this work is a modified version of the filtering proposed in~\cite{baldwin2019inceptive}. In this work, event $t_i(\bm{X})$ is classified into one of the following three categories: Background Activity (BA), Inceptive Event (IE), or Trailing Event (TE). They are defined as:
\begin{align}
    \text{BA}:& \{t_i(\bm{X})-t_{i-1}(\bm{X})>\tau\}\cap  \{t_{i+1}(\bm{X})-t_i(\bm{X})>\tau\}\notag\\
    \text{IE}:& \{t_i(\bm{X})-t_{i-1}(\bm{X})>\tau\}\cap  \{t_{i+1}(\bm{X})-t_i(\bm{X})<\tau\}\notag\\
    \text{TE}:& \{t_i(\bm{X})-t_{i-1}(\bm{X})<\tau\},
\end{align}
where $\tau$ is a threshold value. Intuitively, a single log-intensity change often trigger multiple events of the same polarity in rapid temporal succession. IE corresponds to the first of these events, indicating an arrival of an edge. IE is followed by the TE, which is proportional in number to the magnitude of the log-intensity change that occurred with the inceptive event. Remaining events are called BA, and they are attributed to noise or random activation events. 

In recognition tasks, it was demonstrated empirically that IE was shown to be most useful for describing object shapes~\cite{baldwin2019inceptive}. In our work, however, we are concerned about the negative impact of random activations (BA) on the distance surfaces. Thus we exclude BA from $\Phi$ in \eqref{eq:edge}, as follows:
\begin{align}\label{eq:denoise}
\Phi' =& \left\{\bm{X} \in \mathbb{Z}^2 \left| {\exists i\ni \{t-\Delta t  \leq t_{i}(\bm{X}) < t\}\cap \atop\{\{t_i(\bm{X})-t_{i-1}(\bm{X})<\tau\}\cup \{t_{i+1}(\bm{X})-t_i(\bm{X})<\tau\}\}}\right.\right\}.
\end{align}
The effectiveness of BA exclusion is evident in Figure~\ref{fig:eventGen}. 

In practice, denoising by BA exclusion improves the accuracy of all DVS-based optical flow methods, not limited to DistSurf-OF. The results in Section~\ref{sec:experiments} are shown with and without the same denoising method applied to all optical flow methods.


\begin{figure}
\centering
\includegraphics[width=0.8\linewidth]{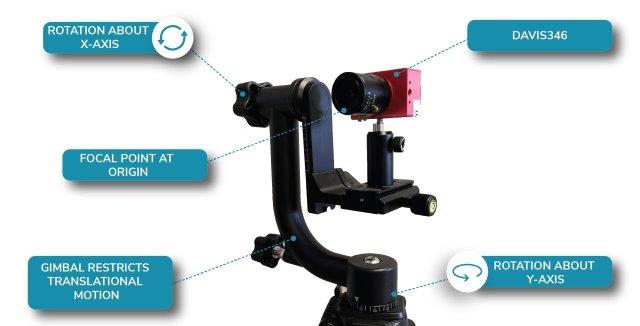}
   \caption{Camera setup for DVSMOTION20 collection. Gimbal limits camera motion while centering the focal point at the origin.} \label{gimbal}
\end{figure}

\begin{figure*}
  \centering
 \begin{tabular}{@{}c@{~}c@{~}c@{~}c@{~}c@{}}
  \includegraphics[width=.25\textwidth]{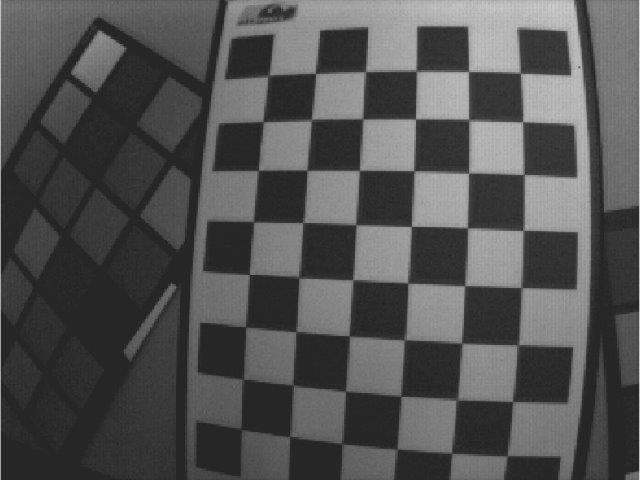}&
     \includegraphics[width=.25\textwidth]{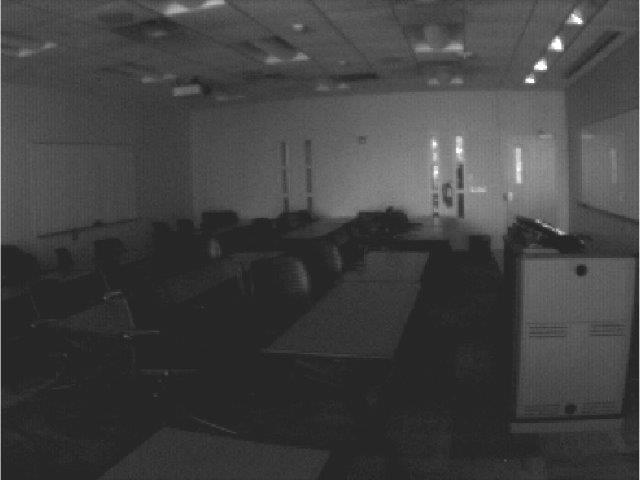}&
    \includegraphics[width=.25\textwidth]{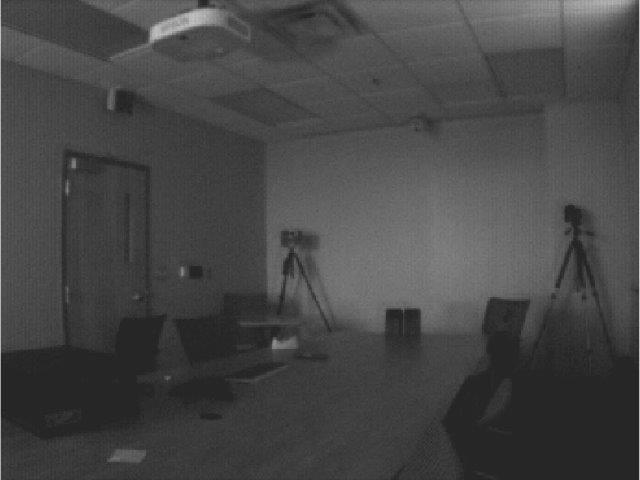}&
     \includegraphics[width=.25\textwidth]{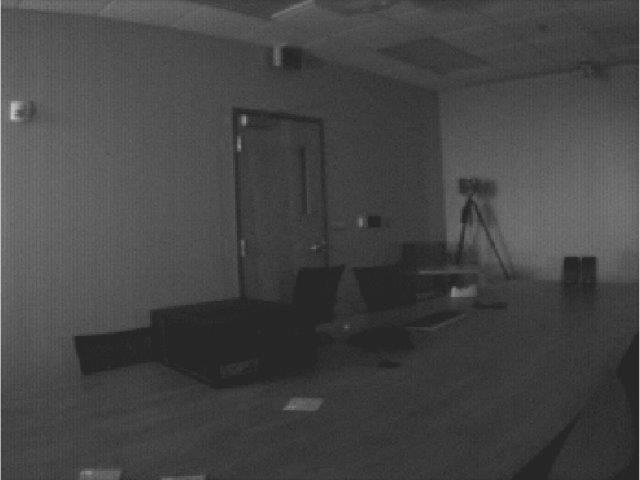}\\
     
      \includegraphics[width=.25\textwidth]{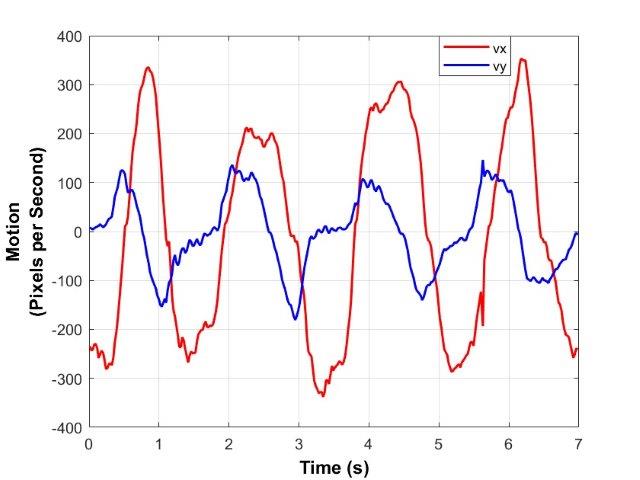}&
     \includegraphics[width=.25\textwidth]{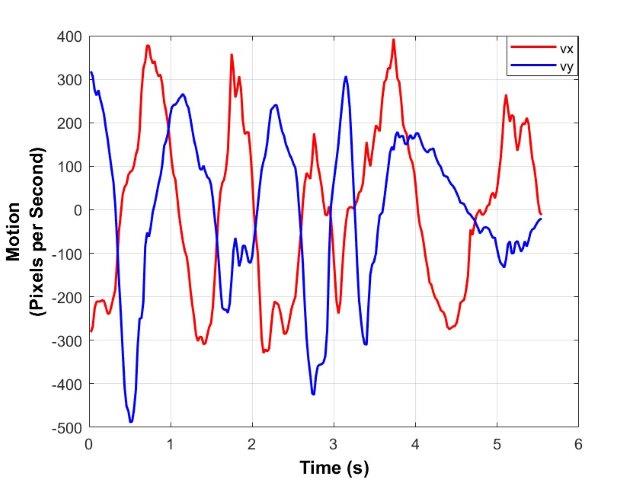}&
    \includegraphics[width=.25\textwidth]{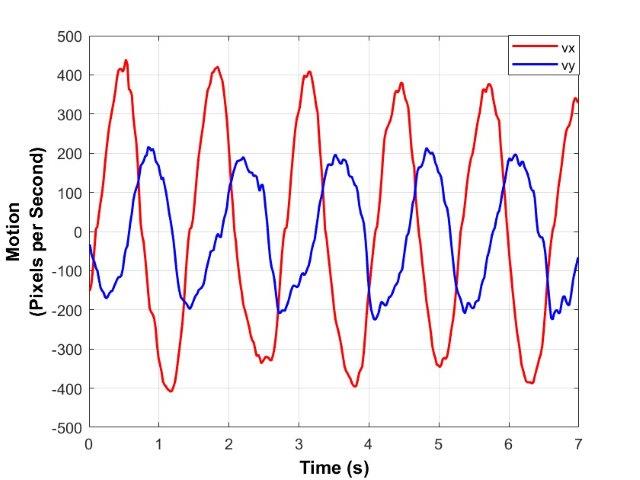}&
     \includegraphics[width=.25\textwidth]{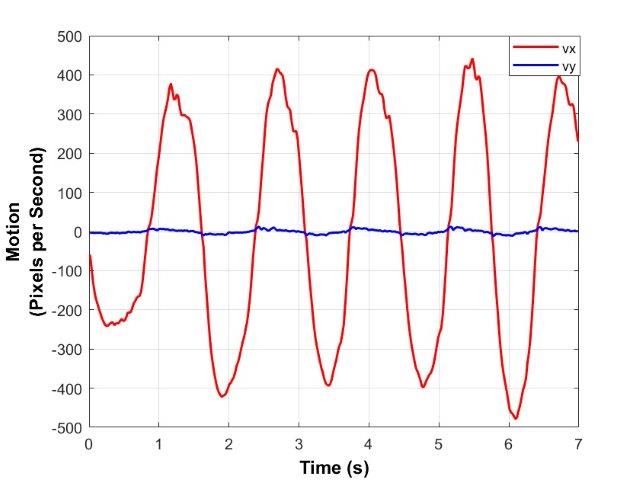}\\

(a) Checkerboard &(b) Classroom  &(c)  Conference Room &(d) Conference Room Translation\\
 \end{tabular}    
\caption{(Top row) Example APS frames extracted from DVSMOTION20 dataset. (Bottom row) IMU measurements recording the trajectory of the camera motion. \label{fig:dataset} }
\end{figure*}

\begin{figure*}
  \centering
 \begin{tabular}{@{}c@{~}c@{~}c@{~}c@{~}c@{}}
 
    \includegraphics[width=.18\textwidth]{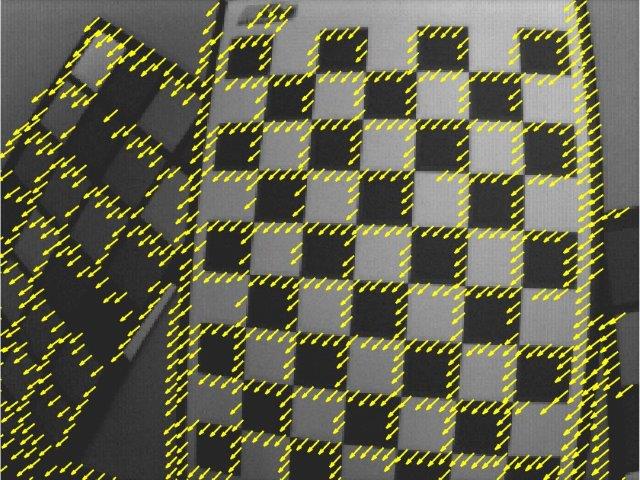}&
    \includegraphics[width=.18\textwidth]{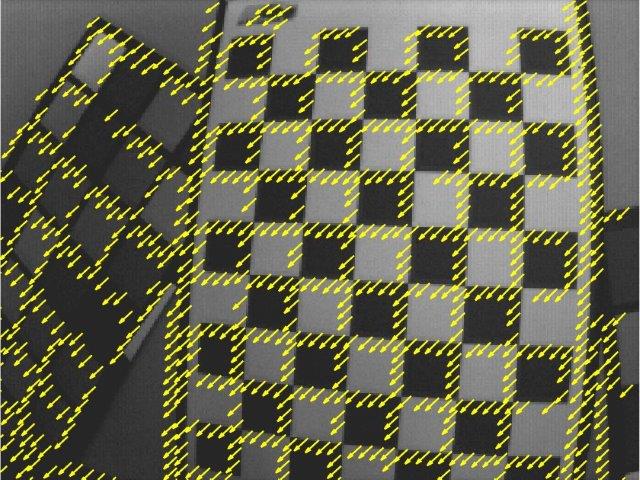}&
     \includegraphics[width=.18\textwidth]{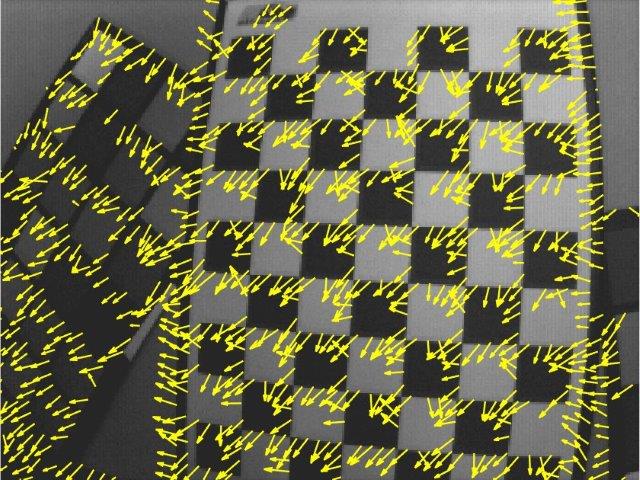}&
     \includegraphics[width=.18\textwidth]{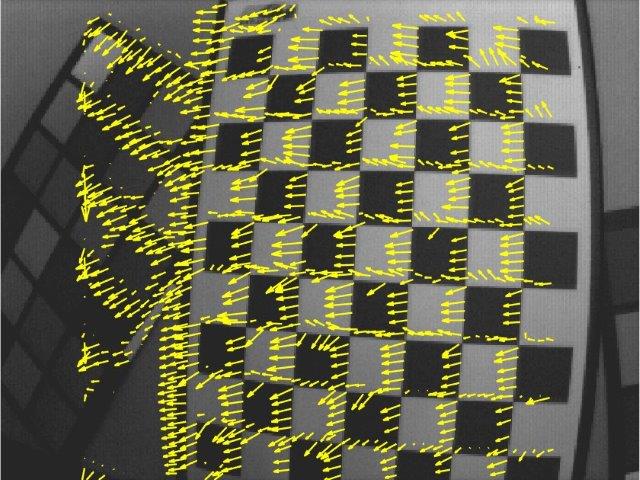}&
    \includegraphics[width=.18\textwidth]{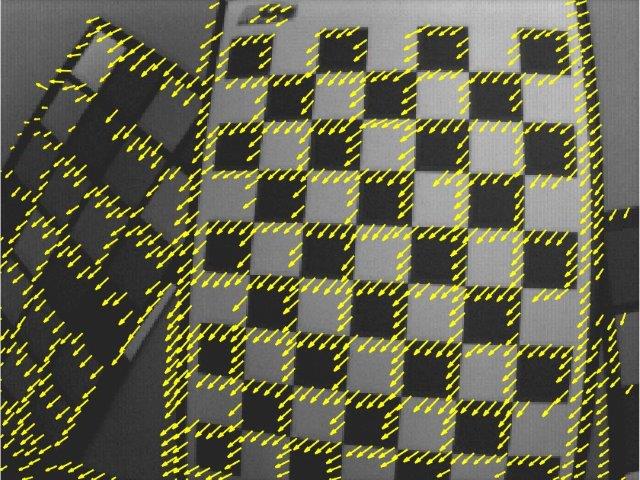}\\

    \includegraphics[width=.18\textwidth]{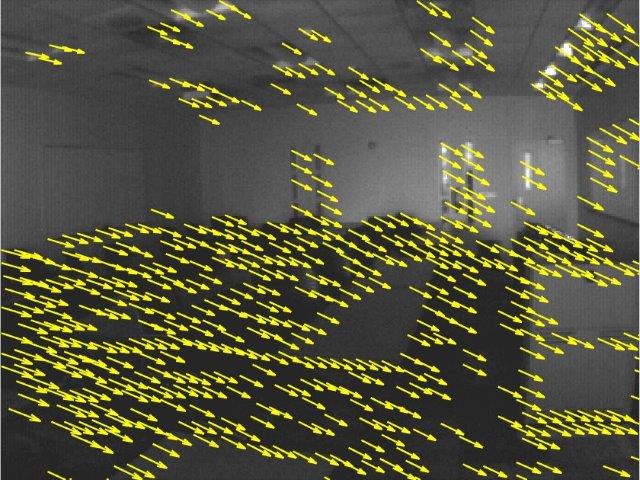}&
    \includegraphics[width=.18\textwidth]{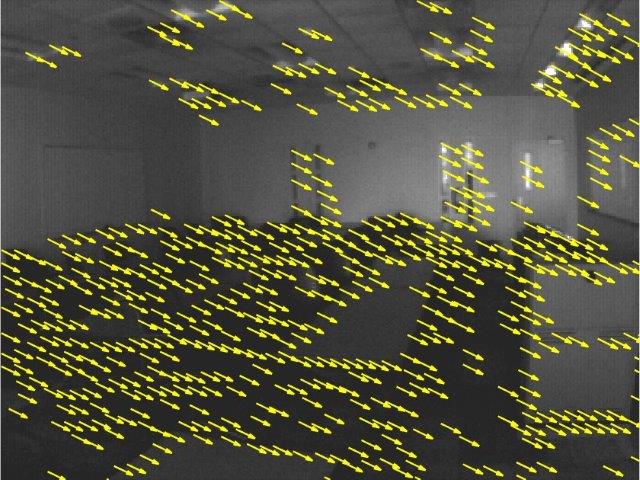}&
    \includegraphics[width=.18\textwidth]{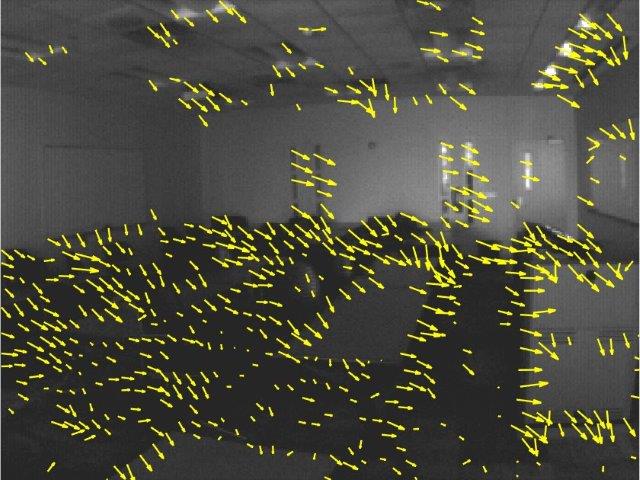}&
     \includegraphics[width=.18\textwidth]{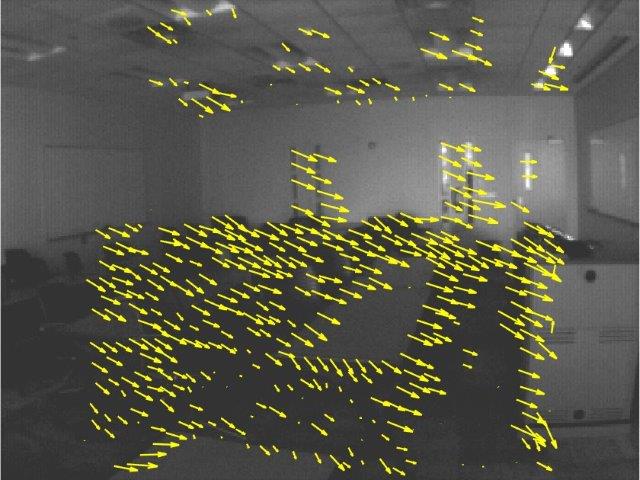}&
    \includegraphics[width=.18\textwidth]{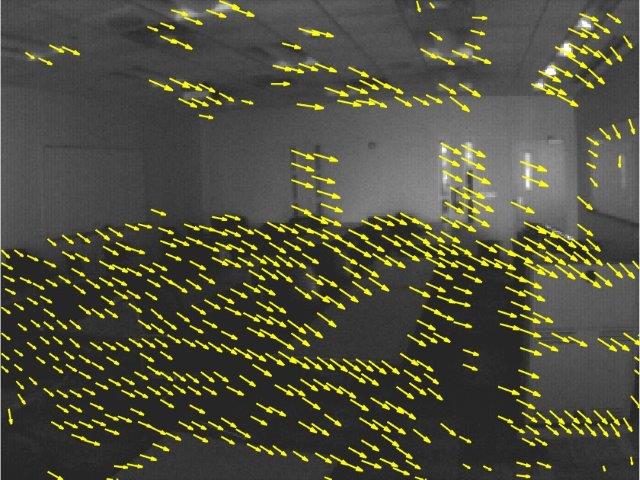}\\

    \includegraphics[width=.18\textwidth]{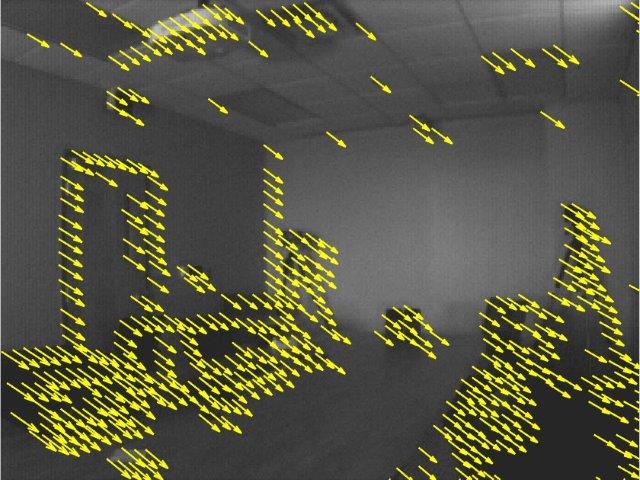}&
    \includegraphics[width=.18\textwidth]{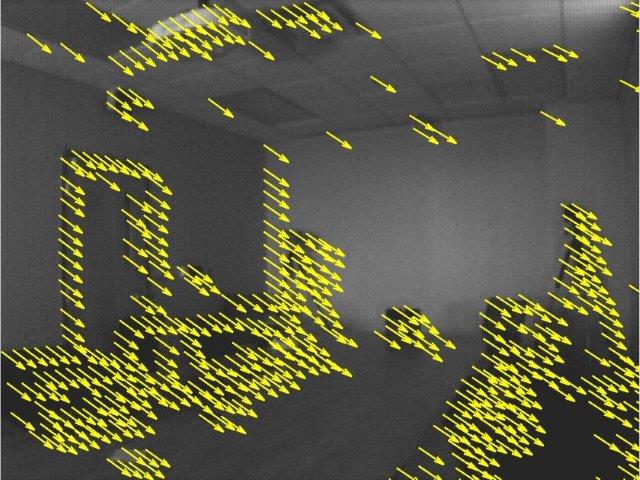}&
    \includegraphics[width=.18\textwidth]{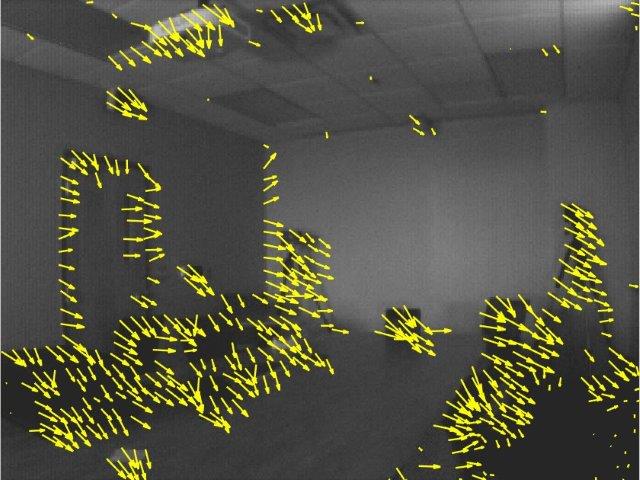}&
    \includegraphics[width=.18\textwidth]{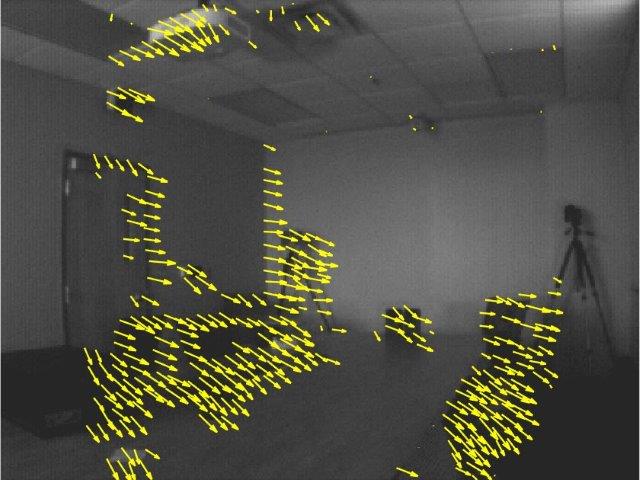}&
    \includegraphics[width=.18\textwidth]{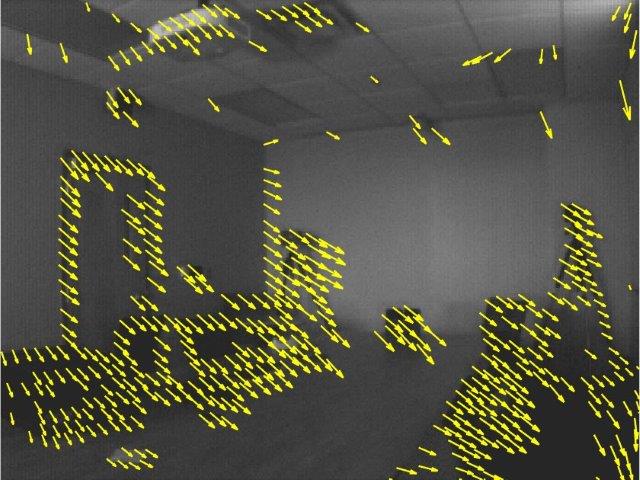}\\
    
    \includegraphics[width=.18\textwidth]{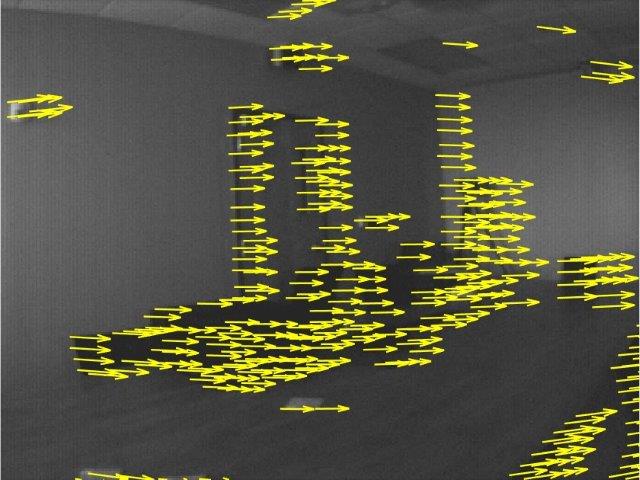}&
    \includegraphics[width=.18\textwidth]{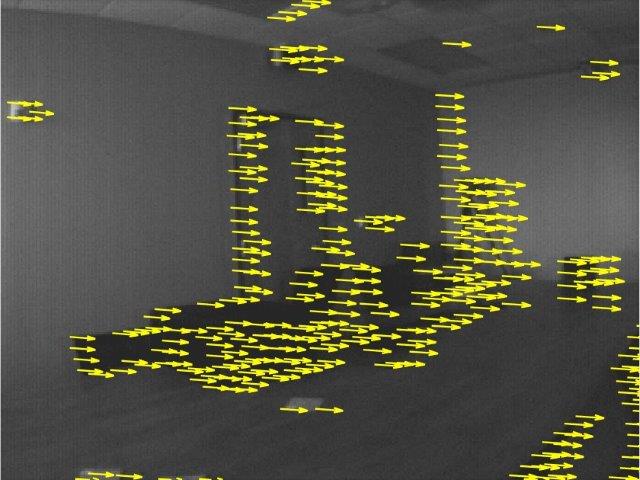}&
    \includegraphics[width=.18\textwidth]{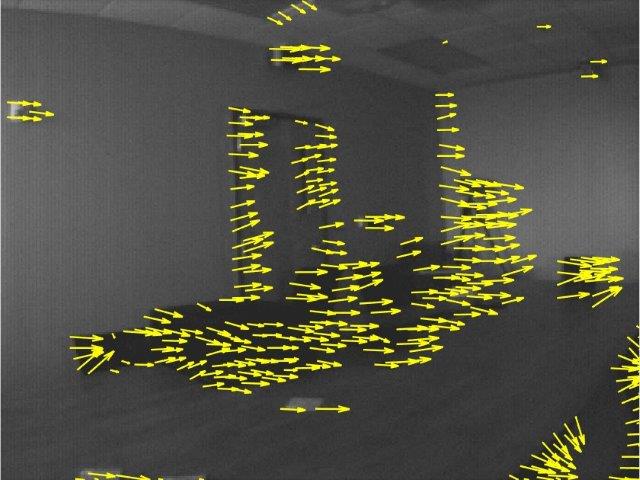}&
    \includegraphics[width=.18\textwidth]{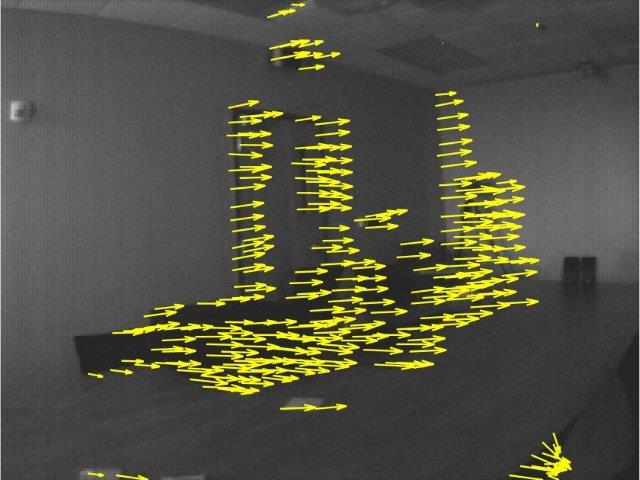}&
    \includegraphics[width=.18\textwidth]{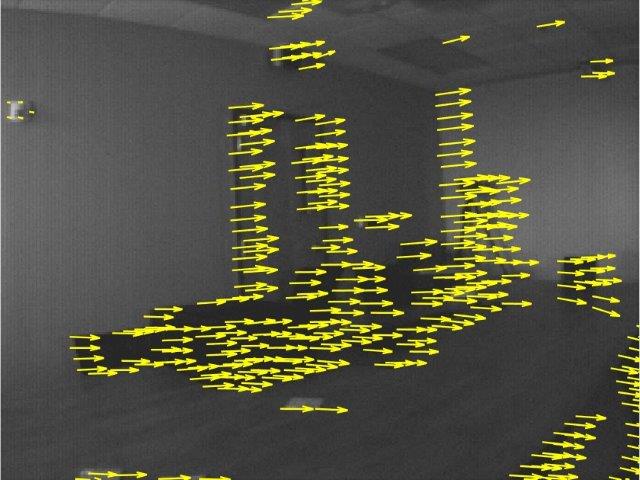}\\

   (a) Ground Truth& (b) DAViS-OF~\cite{almatrafi2019davis} & (c) LK-DVS~\cite{benosman2012asynchronous}  & (d) EV-FlowNet~\cite{zhu2017event}& (e) DistSurf-OF  \\
 \end{tabular}    
\caption{Optical flow results on DVSMOTION20 sequences (shown with denoising). Arrow orientation and magnitude indicate the estimated pixel motion orientation and speed of the observed events. Ground truth is obtained from IMU. See Section~\ref{sec:dataset}. Motion is overlaid on APS for visualization.  \label{OF_res} }
\end{figure*}

\begin{figure*}
  \centering
    \begin{tabular}{@{}c@{~}c@{~}c@{~}c@{~}c@{~}c@{}}
    \includegraphics[width=.18\textwidth]{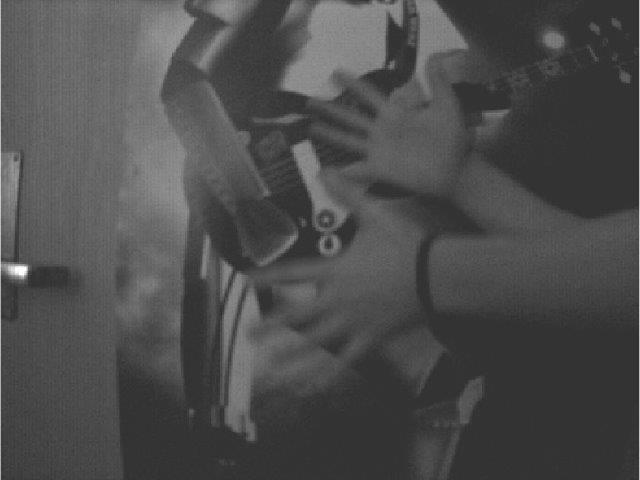}&
    \includegraphics[width=.18\textwidth]{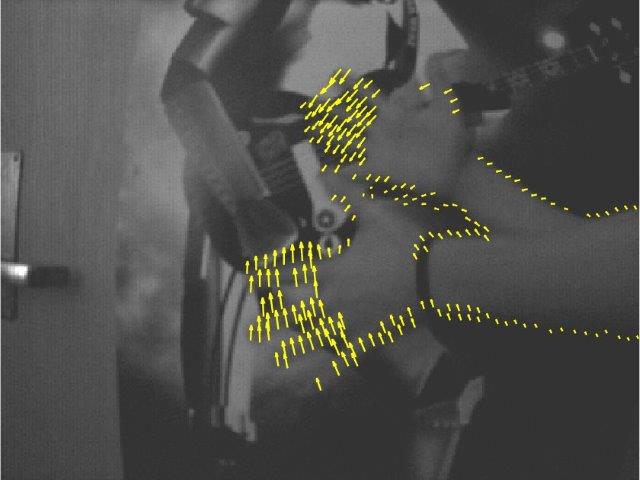}&
    \includegraphics[width=.18\textwidth]{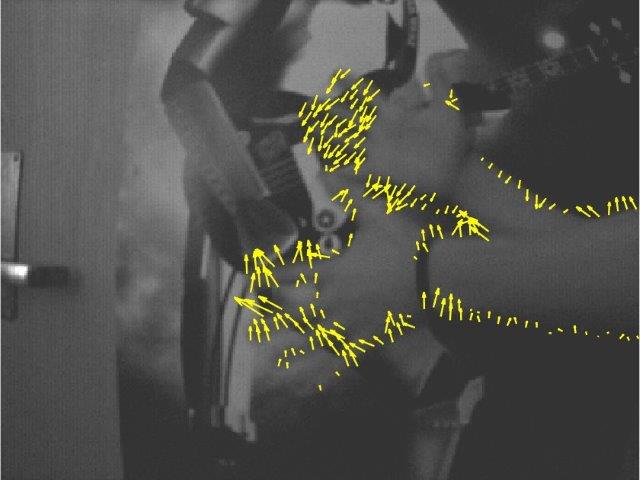}&
     \includegraphics[width=.18\textwidth]{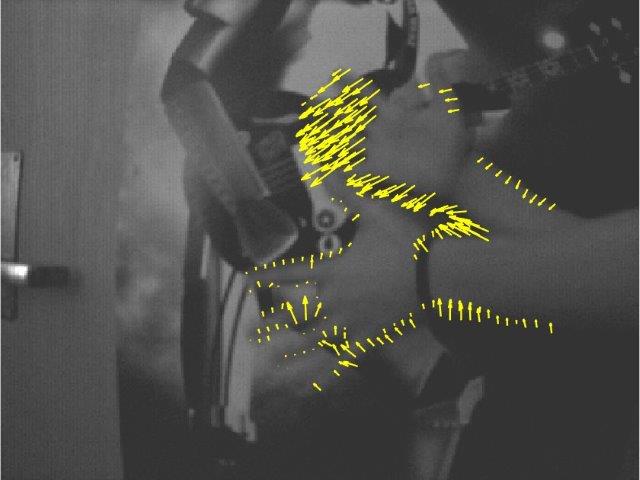}&
    \includegraphics[width=.18\textwidth]{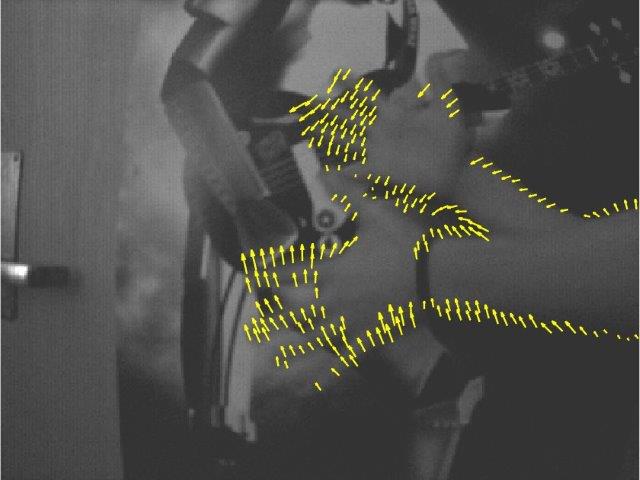}\\

     \includegraphics[width=.18\textwidth]{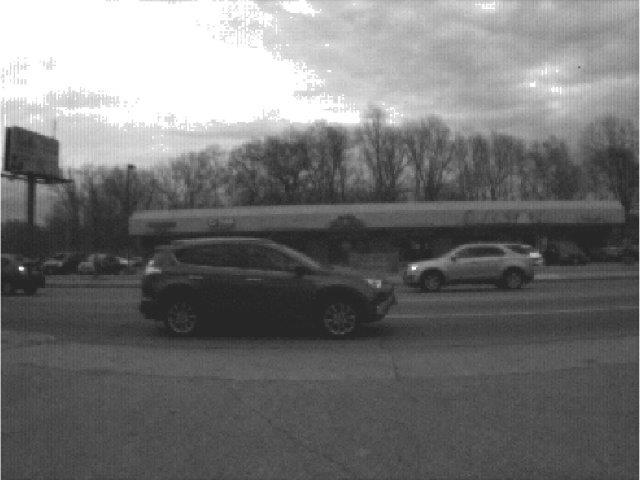}&
     \includegraphics[width=.18\textwidth]{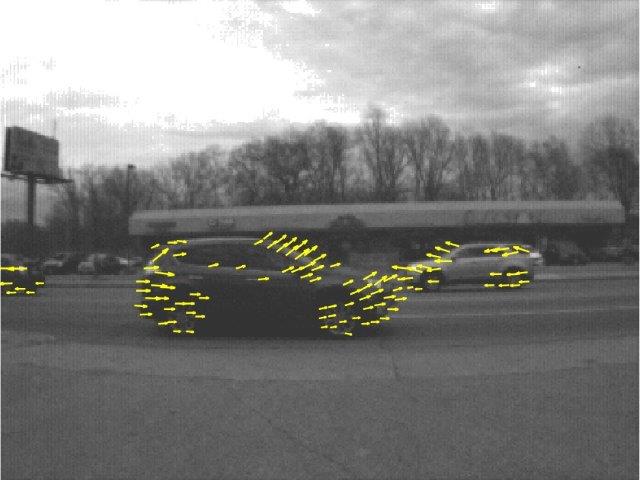}&
     \includegraphics[width=.18\textwidth]{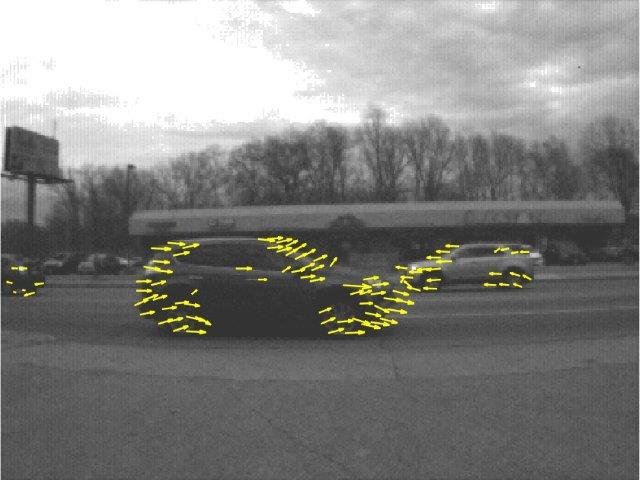}&
     \includegraphics[width=.18\textwidth]{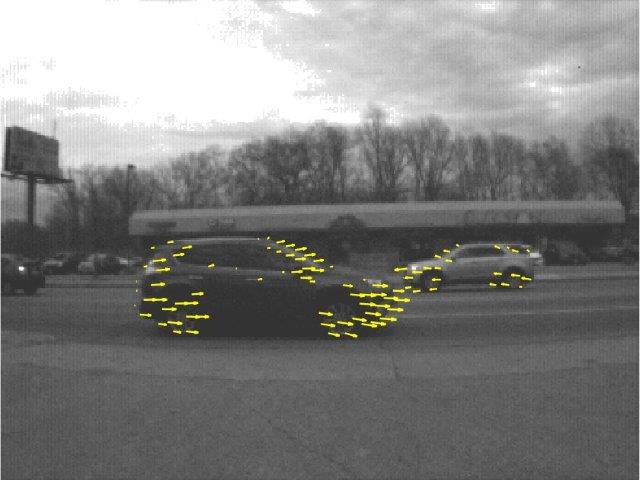}&
  \includegraphics[width=.18\textwidth]{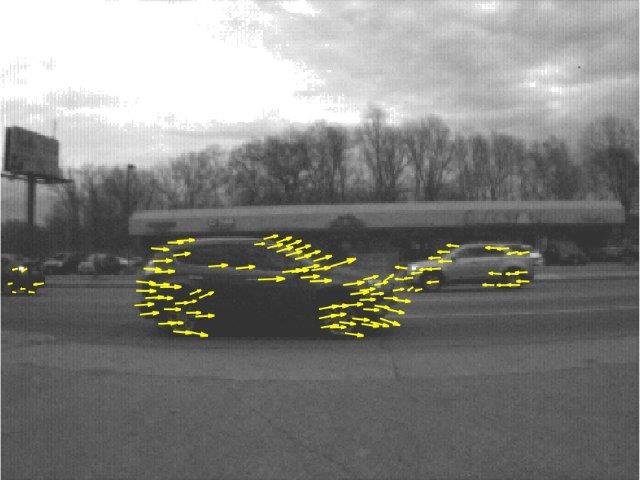}\\
(a)  Frame & (b) DAViS-OF~\cite{almatrafi2019davis} & (c) LK-DVS~\cite{benosman2012asynchronous} & (d) EV-FlowNet~\cite{zhu2017event}&(e) DistSurf-OF
 \end{tabular}    
\caption{Optical flow results on DVSMOTION20 sequences (shown with denoising) containing object motions and motion occlusions. (top row) Hand and (bottom row) car sequences. \label{fig:hand_car} }
\end{figure*}

\begin{table*}
\centering
\caption{Average angular error (AAE) and relative average end-point error (RAEE) with standard deviations. {\bf Bold} font indicates the best DVS-based optical flow performance.}\label{table:testing}
\begin{adjustbox}{width=\textwidth}
\begin{tabular}{l|l|ll|ll|ll|ll} 
  \toprule
  \centering
  {\bf Data Type} & {\bf Method}+{\bf Enhancements} & \multicolumn{2}{c}{{\bf Checkerboard}} &  \multicolumn{2}{c}{\bf Classroom}&\multicolumn{2}{c}{\bf Conference Room}&\multicolumn{2}{c}{\bf Conference Room Translation}\\
   &&{\bf RAEE [\%] }& {\bf AAE [\textdegree]} & {\bf RAEE [\%]} & {\bf AAE [\textdegree]}& {\bf RAEE [\%]} & {\bf AAE [\textdegree]}& {\bf RAEE [\%]} & {\bf AAE [\textdegree]}\\
  \midrule \midrule
&LK-DVS \cite{benosman2012asynchronous} + \cite{event2016dataset}&$39.30\pm21.4$&$13.19\pm14.40$&$53.42\pm23.13$&$20.45\pm20.68$&$47.39\pm25.43$&$21.49\pm26.41$&$48.21\pm25.87$&$21.19\pm28.88$\\
DVS
    &LP \cite{benosman2014event} + \cite{barranco2014contour,event2016dataset}&$121.66\pm35.35$&$91.00\pm31.03$&$110.76\pm40.56$&$92.61\pm50.24$&$117.45\pm42.68$&$98.10\pm50.40$&$117.10\pm143.71$&$81.79\pm47.76$\\ 
    & EV-FlowNet \cite{zhu2018ev}&$36.12\pm23.67$&$15.86\pm16.61$&$ 53.42\pm45.92$&$18.77\pm14.33$&$41.24\pm31.49$&$18.37\pm18.29$& $28.37\pm25.00$&$10.63\pm14.56$ \\
  \midrule
  &LK-DVS \cite{benosman2012asynchronous} + \cite{event2016dataset}&$42.32\pm20.53$&$12.06\pm11.15$&$32.64\pm17.29$&$11.66\pm11.48$&$30.81\pm16.91$&$11.87\pm11.37$&$31.48\pm16.81$&$10.97\pm11.46$\\
DVS + &LP \cite{benosman2014event} + \cite{barranco2014contour,event2016dataset}&$104.22\pm17.06$&$72.07\pm18.23$&$103.09\pm31.79$&$82.96\pm42.53$&$102.08\pm28.67$&$76.35\pm40.82$&$112.25\pm111.23$&$78.23\pm40.20$\\ 
denoising&EV-FlowNet\cite{zhu2018ev}&$38.56\pm24.03$&$16.97\pm16.17$& $44.62\pm24.18$&$15.14\pm16.77$&$38.52\pm22.21$&$16.27\pm15.92$& $23.62\pm20.06$&$8.04\pm11.82$\\ 
   &DistSurf-OF (proposed)&${\bf 20.54}\pm10.36$&${\bf 5.89}\pm5.11$&${\bf 25.62}\pm13.27$&${\bf 6.84}\pm6.26$&${\bf 18.71}\pm12.38$&${\bf 5.67}\pm5.75$&${\bf 18.71}\pm12.95$&${\bf 5.42}\pm5.38$\\


   \midrule
   
 {DAViS}&DAViS-OF \cite{almatrafi2019davis}&$17.56\pm8.18$&$3.80\pm2.64$&$20.00\pm6.60$&$5.01\pm2.69$&$19.77\pm6.32$&$5.48\pm2.71$&$23.76\pm5.74$&$3.77\pm1.97$ \\
  \bottomrule
\end{tabular}
\end{adjustbox}
\label{res:table}
\end{table*}

\section{Experimental Result} \label{sec:experiments}

\subsection{DVSMOTION20 Dataset}
\label{sec:dataset}
The existing benchmarking datasets have played a critical role in the progress of research in optical flow for neuromorphic cameras\cite{zhu2018ev,rueckauer2016evaluation,almatrafi2019davis}. However, they are not without shortcomings. Motion and the scene contents in some sequences of \cite{rueckauer2016evaluation} are overly short, simplistic, and unnaturally favor normal flow (where the motion is perpendicular to the edge orientation). The dataset in \cite{zhu2018ev} update ground truth motion at 20Hz sampling rate---slow considering that DVS is accurate to microseconds. The spatial resolution of sequences in \cite{almatrafi2019davis} is small.

Therefore, we collected a new dataset---called DVSMOTION20---using IniVation DAViS346 camera in attempt to further enhance the progress of DVS-based optical flow methods. The DAViS346 camera has a 346$\times$260 spatial resolution and outputs frames (APS) up to 60 frames per seconds, events in microsecond resolution, and a 6-axis IMU data at around 1kHz sampling rate. We used a standard checkerboard calibration target to recover the intrinsic parameters of the camera. We infer ground truth pixel velocity stemming from camera motion using the inertial measurement unit (IMU), similar to prior benchmarkings in \cite{rueckauer2016evaluation,almatrafi2019davis}. Specifically, we placed the camera on a gimbal as shown in  Figure~\ref{gimbal}, restricting the movement to yaw, pitch, and roll rotations (i.e.~no translations). This restriction to angular rotational motion ensures that pixel velocity can be recovered entirely from gyroscope data.

DVSMOTION20 dataset contains four real indoor sequences (\emph{checkerboard}, \emph{classroom}, \emph{conference room}, and \emph{conference room translation}). Each scene was captured for around 13-16 seconds with the first three seconds containing no motion for IMU calibration; 7-8 seconds of DVS data following the IMU calibration is used in the performance evaluation in Section~\ref{sec:results}. {Each recorded data files are about 500MB in size.} See Figure~\ref{fig:dataset} for example frame content and the IMU trajectories. Although the sequences were restricted to camera motion with a stationary scene, all except for \emph{conference room translation} contain fast motion with complex random camera movements---they should be challenging to optical flow methods that tend to yield normal flow. In \emph{conference room translation} the motion was purely horizontal (yaw rotation). It helps verify the hypothesis that most optical flow methods generally work better when the pixel motion is simple.

We provide additional sequences in DVSMOTION20 (called \emph{hands} and \emph{cars}) containing multiple object motions (i.e.~not camera motion). Object motion is spatially local by nature, but with strong motion boundaries. In our sequences, cars and hands moving in the opposite direction intersect, resulting in motion occlusion. See Figure~\ref{fig:hand_car}. Visual inspection of the estimated motion vectors are more than adequate for showing optical flow failures in presence of occlusion or large object motion (which happens frequently), even if the ground truth motion vectors are not available (because object motion cannot be inferred from IMU).

\subsection{Setup and Comparisons}

We compared DistSurf-OF to several state-of-the-art DVS-based optical flow algorithms including LK-DVS~\cite{benosman2012asynchronous}, LP~\cite{benosman2014event}, and EV-FlowNet~\cite{zhu2018ev}. In particular, the learning-based method in \cite{zhu2018ev} (trained originally using DVSFLOW16) is fine tuned with sequences in DVSMOTION20. Since training and testing data are identical, results we show in this paper represents its \emph{best case scenario} performance.

We also compared to the state-of-the-art DAVIS optical flow method DAVIS-OF~\cite{almatrafi2019davis}. Because this method makes use of DVS (events, high temporal fidelity) as well as APS (intensity, high spatial fidelity) data, we expect it to perform better than the DVS-only methods. As evidenced in Table~\ref{table:testing}, however, the performance of proposed DistSurf-OF comes surprisingly close to this DAVIS method in some sequences.

To ensure a fair evaluation, we applied the same $\Delta t=$5ms to all methods using the temporal window in the evaluation. That is, we output a motion field every 5ms. We also found empirically that the denoising technique in \eqref{eq:denoise} improved the performance of \emph{all} DVS-based optical flow methods (not just ours). Thus, the results shown in Section~\ref{sec:results} are shown with and without the same denoising for all optical flow methods. 

The proposed DistSurf-OF method was implemented and run on MATLAB 2019b operating on Lenovo ThinkStation P520C. In our implementation, we fixed the temporal window $\Delta t$ used for edge map in \eqref{eq:edge} and the predefined threshold parameter $\tau$ for denoising in \eqref{eq:denoise} to 5ms. The spatial gradient filter used in \eqref{eq:spatialgradient} was {the 4-point central difference (with mask coefficients ($\frac{1}{12}(-1,8,0,-8,1)$)~\cite{barron1994performance}}. With these configurations, the execution time to yield a motion field for the 346$\times$260 spatial resolution sensor was around 0.737s. However, the computation of operations specific to DistSurf-OF (event denoising, temporal windowing, distance transform, and spatial/temporal derivatives) take only 35ms, while the (intensity-based) optical flow method in \cite{sun2014quantitative} takes 0.702s. Therefore, users have the freedom to choose different intensity-based optical flow methods to pair with DistSurf based on the accuracy and speed requirements of the application. The computation of DistSurf-OF itself can be sped up further by GPU-based parallel coding, as well as fast distance transform implementations (e.g.~\cite{karam2019fast}) that reduce complexity from $\mathcal{O}(n^2)$ to $\mathcal{O}(n\log{}n)$ or even $\mathcal{O}(n)$. The code and the DVSMOTION20 dataset (with the accompanying ground truth motion field) are made available at \url{issl.udayton.edu}.

\subsection{Results and Discussion}
\label{sec:results}

The performances of DistSurf-OF and the state-of-the-art DVS/DAVIS optical flow methods on sequences shown in Figure~\ref{OF_res} are reported in Table~\ref{res:table}. For error statistics, we show average angular error (AAE) and the relative average end-point error (RAEE), referring to the pixel motion magnitude and angle errors, respectively\cite{rueckauer2016evaluation} and \cite{almatrafi2019davis}. The stability of each method can be inferred from the standard deviations of the error statistics.

Among the DVS-based optical flow methods, DistSurf-OF has a clear advantage. Average angular error is consistently below 7\textdegree~in all sequences, while the motion magnitude as evaluated by AEE is also the smallest. Contrast this to the other state-of-the-art DVS-based optical flow methods in \cite{benosman2012asynchronous,benosman2014event,delbruck2008frame}, whose angular error exceeds 10\textdegree~in all but simplistic motion sequence (\emph{translation conference}). Method in \cite{benosman2014event} failed in three of the four sequences. One can also gauge the effectiveness of the denoising on the state-of-the-art methods by comparing the ``DVS'' and ``DVS+denoising'' data types in Table~\ref{table:testing}. In all but the \emph{checkerboard} sequence, optical flow applied to IE+TE in \eqref{eq:denoise} was more accurate than same methods applied to IE+TE+BA.

The performance of DistSurf-OF is closer to that of the state-of-the-art DAViS-OF method in \cite{almatrafi2019davis}. Owing to the fact that the latter method leverages both DVS and APS data, it is indeed expected to perform better than DVS-only methods. With an angular error around 5\textdegree, there are about 2\textdegree~only separating the performance of DistSurf-OF (DVS only) and DAViS-OF (DVS+APS).

The trends in Table~\ref{res:table} can be visually confirmed by the results in Figure~\ref{OF_res}. DistSurf-OF in Figure~\ref{OF_res}(c) yields stable and satisfactory results in terms of motion orientation and magnitude, closely resembling the ground truth motion field in Figure~\ref{OF_res}(a) and DAViS-OF in Figure~\ref{OF_res}(b). In particular, the motion detected by the proposed method in \emph{checkerboard} sequence is in the correct orientation, not normal to the edge direction. Contrast this to EV-FlowNet in Figure~\ref{OF_res}(d), whose estimated motion direction is predominantly horizontal (perpendicular to the vertical edges). LK-DVS in Figure~\ref{OF_res}(c) is more accurate than EV-FlowNet in terms of motion orientation, but lacks spatial consistency.

The other sequences are rich with diverse edge orientations and edge length, making it possible to assess the optical flow method's robustness to real-world variations. DistSurf-OF handled them well, save for the events occurring very close to the boundaries of the sensor. The quality of the estimated motion field is comparable to that of the DAVIS-OF. EV-Flownet performed better in the \emph{checkerboard} sequence, although the proposed method was still better when comparing the orientations of the pixel motions event-for-event to the ground truth motion. LK-DVS is unable to resolve the spatial inconsistency problem.

Finally, Figure~\ref{fig:hand_car} shows more challenging sequences containing multiple objects moving in opposite directions. In the \emph{hand} sequence where two hands cross in front of a textured background, DAVIS-OF and DistSurf-OF are able to track individual fingers and their corresponding directions accurately, and the motion boundaries where the hands and the arms cross each other is well defined. Despite the lack of spatial consistency, LK-DVS largely detects the orientation of the finger movement. The high concentration of events in the fingers seem to confuse the pixel motion estimation in EV-FlowNet, with inconsistent velocity magnitudes and orientations.

In the \emph{car} sequence, the motion estimated by all methods on the windshield of the foreground car seem to point towards the sky erroneously---meaning all optical flow methods yielded normal flow. In the remainder of the foreground car and the background, DistSurf-OF's motions are horizontally oriented with consistent motion magnitudes. EV-FlowNet estimates of the motion orientation in the remainder of foreground car are more stable, but the motion magnitudes vary considerably; and the estimated background car motion is inconsistent with the context. The quality of LK-DVS is comparable to DistSurf-OF.

\section{Conclusion} \label{sec:conclusion}
We proposed the notion of \emph{distance surface} for performing optical flow tasks in neuromorphic cameras. We proposed to use the distance transforms computed from the events generated from DVS as a proxy for object textures. We rigorously proved that distance surface satisfy optical flow equations, and the event pixel motion recovered by DistSurf-OF are highly accurate. We verified the effectiveness of our method using DVSMOTION20 dataset. For future work, we plan to investigate whether DistSurf can be combined with APS (similar in style to \cite{almatrafi2019davis}) to further improve the optical flow accuracy.

\ifpeerreview \else
\section*{Acknowledgments}
This work was made possible in part by funding from Ford University Research Program and the Japan National Institute of Information and Communications Technology.
\fi

\bibliographystyle{IEEEtran}
\bibliography{references}

\ifpeerreview \else






\begin{IEEEbiography}[{\includegraphics[width=1in,height=1.25in,clip,keepaspectratio]{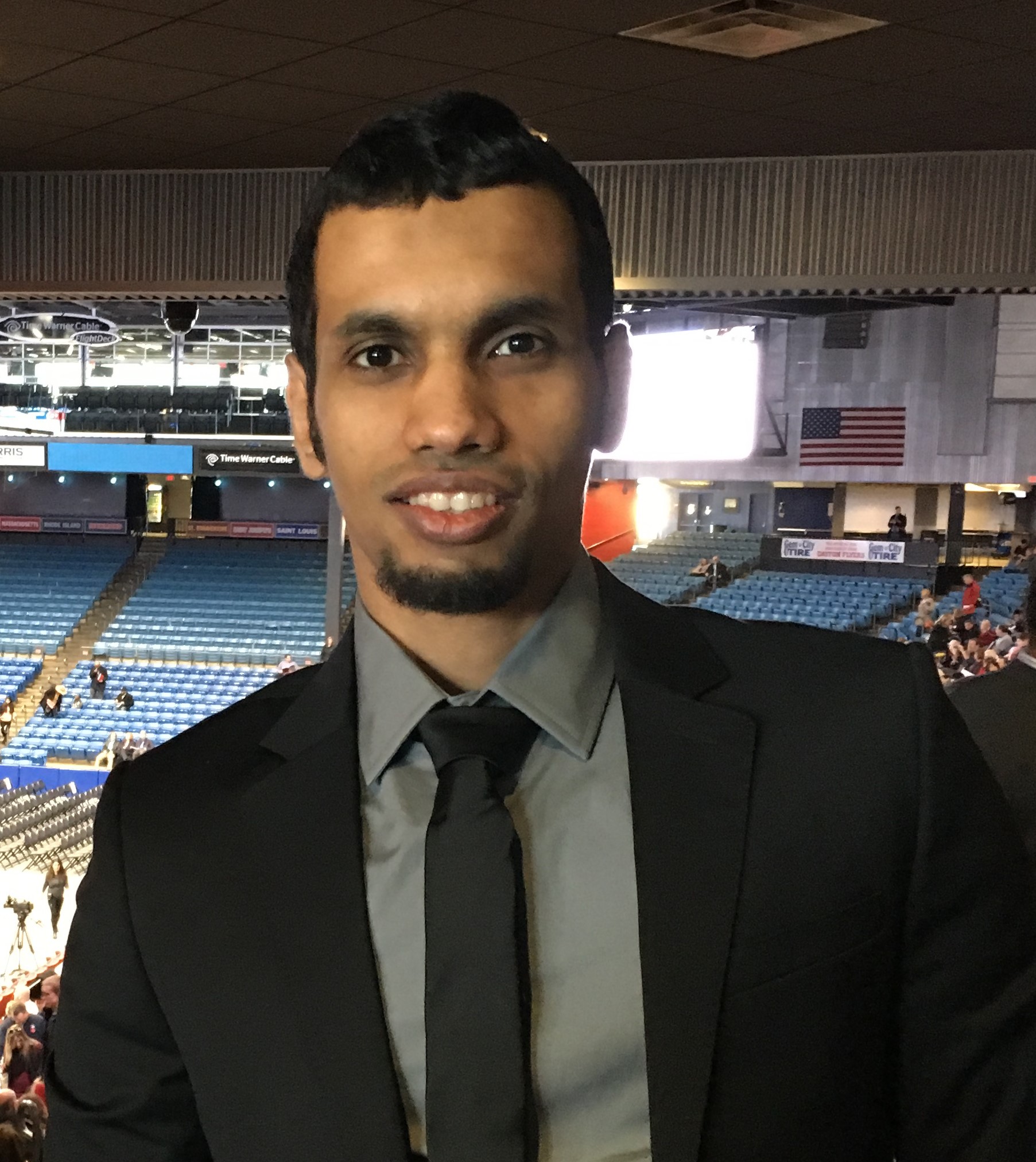}}]{Mohammed Almatrafi} 
(S’15-M’20) received his B.S. (Hons.) in Electrical Engineering from Umm Al-Qura University, Makkah, Saudi Arabia in 2011, the M.S. and Ph.D. degrees in Electrical and Computer Engineering from University of Dayton, Dayton, OH, USA, in 2015 and 2019, respectively. He is currently an Assistant Professor and the Vice Dean of the College of Engineering for Postgraduate Studies and Scientific Research at Umm Al-Qura University, Al-Lith, Saudi Arabia. His research is focused on  image processing, neuromorphic cameras and computer vision.   
\end{IEEEbiography}

\begin{IEEEbiography}[{\includegraphics[width=1in,height=1.25in,clip,keepaspectratio]{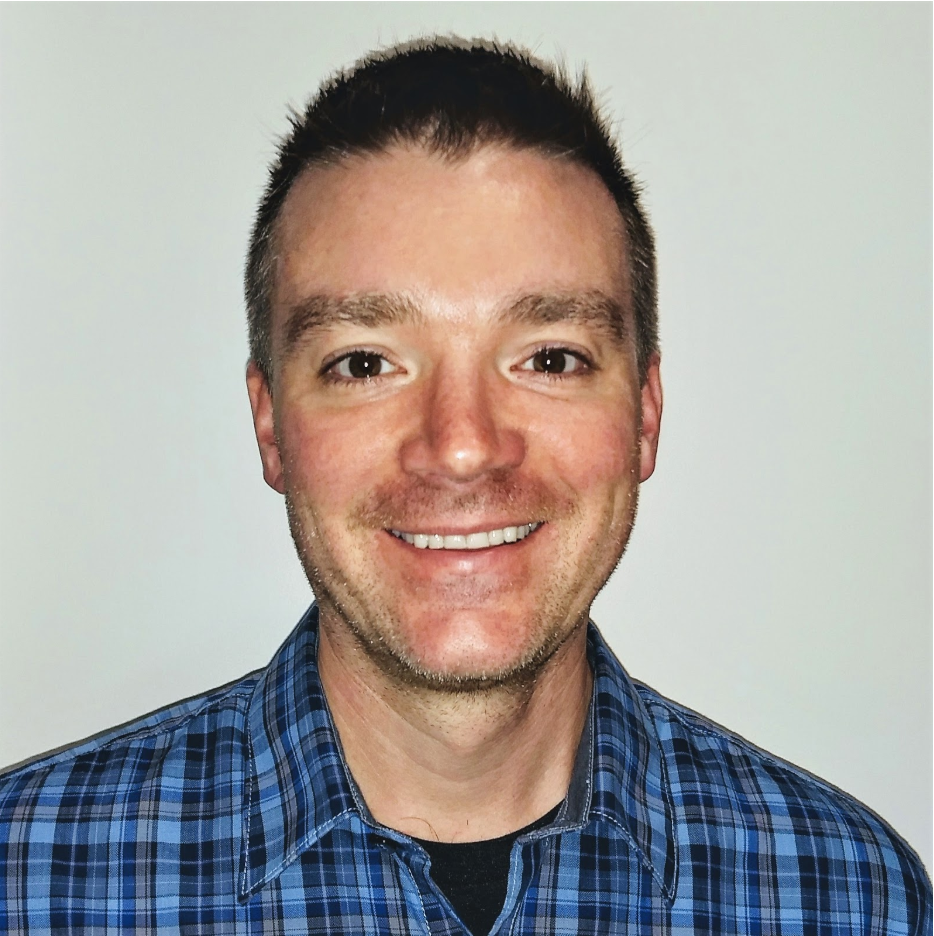}}]{Wes Baldwin}
 received his B.S. in Computer Engineering in 2002 from Kettering University and his M.S. in Electrical and Computer Engineering in 2005 from the University of Illinois at Chicago. He is currently working towards a Ph.D. in Electrical and Computer Engineering at the University of Dayton. His research is focused on machine learning, image processing, and neuromorphic cameras.
\end{IEEEbiography}

\begin{IEEEbiography}[{\includegraphics[width=1in,height=1.25in,clip,keepaspectratio]{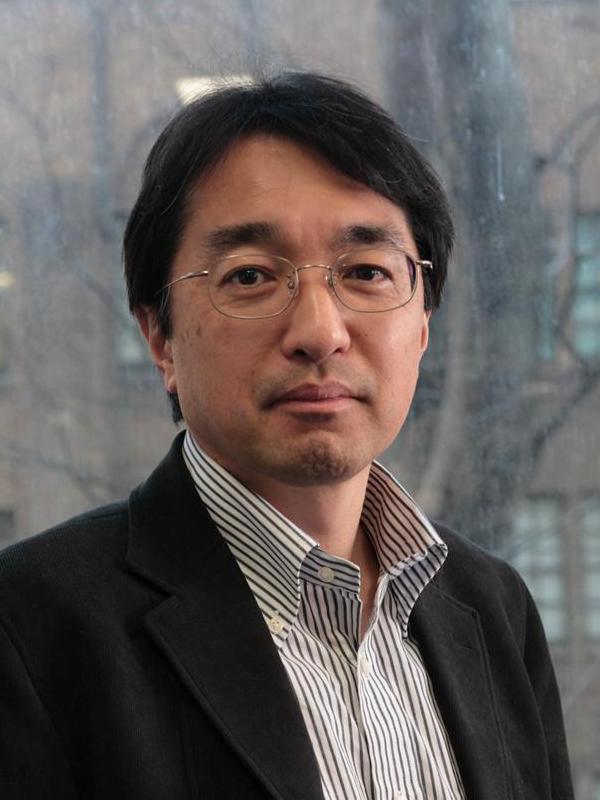}}]{Kiyoharu Aizawa}
received the B.E., the M.E., and the Dr.Eng. degrees in Electrical Engineering
all from the University of Tokyo, in 1983, 1985, 1988, respectively. He is
currently a Professor at Department of Information and Communication
Engineering of the University of Tokyo. He was a Visiting Assistant Professor 
at University of Illinois from 1990 to 1992. He received the 1987 Young Engineer Award and the 1990, 1998 Best Paper Awards, 
the 1991 Achievement Award, 1999 Electronics Society Award from IEICE Japan, 
and the 1998 Fujio Frontier Award, the 2002 and 2009 Best Paper Award, and  
2013 Achievement award from ITE Japan. He received the IBM Japan Science Prize in 2002.He is on Editorial Boards of IEEE MultiMedia, ACM TOMM,
APSIPA Transactions on Signal and Information Processing, and  
International Journal of Multimedia Information Retrieval.
He served as the Editor in Chief of Journal of ITE Japan,
an Associate Editor of IEEE Trans. Image Processing, IEEE Trans. CSVT and IEEE Trans.
Multimedia. He is/was a president of ITE and ISS society of IEICE, 2019 and 2018, respectively.
He has served a number of international and domestic
conferences; he was a General co-Chair of ACM Multimedia 2012 and ACM ICMR2018. 
He is a Fellow of IEEE, IEICE, ITE and a council member of Science Council of Japan. His research interest is in multimedia applications, image processing and computer vision
\end{IEEEbiography}

\begin{IEEEbiography}[{\includegraphics[width=1in,height=1.25in,clip,keepaspectratio]{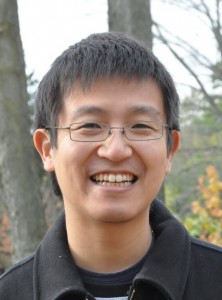}}]{Keigo Hirakawa}
(S’00–M’05–SM’11) received the
B.S. degree (Hons.) in electrical engineering from
Princeton University, Princeton, NJ, USA, in 2000,
the M.S. and Ph.D. degrees in electrical and computer engineering from Cornell University, Ithaca,
NY, USA, in 2003 and 2005, respectively, and the
M.M. degree (Hons.) in jazz performance studies
from the New England Conservatory of Music,
Boston, MA, USA, in 2006. He was a Research
Associate with Harvard University, Cambridge, MA,
USA from 2006 to 2009. He is currently an Associate Professor with the University of Dayton, Dayton, OH, USA. He is
currently the Head of the Intelligent Signal Systems Laboratory, University of
Dayton, where the group focuses on statistical signal processing, color image
processing, and computer vision.
\end{IEEEbiography}

\fi

\end{document}